\documentclass[article]{article}
\pdfoutput=1
\usepackage{jheppub}
\usepackage[utf8]{inputenc}
\usepackage[english]{babel}

\def\bi{\begin{itemize}}
	\def\ei{\end{itemize}}
 \def\be{\begin{equation}}
\def\ee{\end{equation}}
 \def\ba{\begin{align}}
\def\ea{\end{align}}
\def\bea{\begin{eqnarray}}
\def\eea{\end{eqnarray}}
\def\pd{\partial}
\def\a{\alpha}
\def\b{\beta}
\def\g{\gamma}
\def\d{\delta}
\def\m{\mu}
\def\n{\nu}
\def\t{\tau}

\def\l{\lambda}
\def\O{\Omega}
\def\r{\rho}
\def\s{\sigma}

\def\D{\nabla}

\def\f{\phi}

\def\bg{\bar{g}}
\def\bR{\bar{R}}
\def\bD{\bar{\nabla}}

\preprint{ IFT-UAM/CSIC-15-040 FTUAM-15-12 FTI/UCM 47/2015}
\title{Quantum Corrections to Unimodular Gravity
}
\author{\begin{flushright}\it{ \large\fontfamily{pzc}\selectfont To the memory of Raymond Stora, teacher and friend.}\end{flushright}}
\author[a,b]{\par Enrique Álvarez,}
\author[a,b]{Sergio González-Martín,}
\author[a,b]{Mario Herrero-Valea}
\author[c]{and Carmelo P. Martín}

\affiliation[a]{Instituto de Física Teórica UAM/CSIC\\C/ Nicolas Cabrera, 13–15, C.U. Cantoblanco, 28049 Madrid, Spain}
\affiliation[b]{Departamento de Física Teórica\\Universidad Autónoma de Madrid, 20849 Madrid, Spain}
\affiliation[c]{Universidad Complutense de Madrid (UCM), Departamento de Física Téorica I\\ Facultad de Ciencias Físicas, Av. Complutense S/N (Ciudad Univ.), 28040 Madrid, Spain}

\emailAdd{enrique.alvarez@uam.es, sergio.gonzalez.martin@csic.es,\\ mario.herrero@csic.es, carmelop@fis.ucm.es}

\abstract{The problem of the  cosmological constant appears in a new light in Unimodular Gravity. In particular, the zero momentum piece of the potential (that is, the constant piece independent of the matter fields) does not automatically produce a cosmological constant proportional to it. The aim of this paper is to give some details on a calculation showing that quantum corrections do not renormalize the classical value of this observable.  

}
\begin{document}
 \maketitle
\newpage

\newpage\section{Introduction}

Even if some sort of {\em quintessence model} of dark energy is able to explain the observed acceleration of the Universe it remains another problem. This problem is theoretically independent of the dark energy one, and it consists in understanding why the vacuum energy do not produce a huge value for the cosmological constant, many orders of magnitude above the observed value. While a precise prediction is not available, the order of magnitude seems clear. It would appear that either Wilsonian ideas of effective theories do not work in this case, or else that vacuum energy does not obey the equivalence principle.
\par
In Unimodular Gravity the vacuum energy (actually all potential energy) is naively decoupled from gravitation, because when the spacetime metric is unimodular, that is
\be
\tilde{g}\equiv \text{det}~\tilde{g}_{\m\n}=-1
\ee
the interaction term between the potential energy and the metric is of the type
\be
S_\text{in}\equiv \int d^n x \left({1\over 2}\tilde{g}^{\m\n}\pd_\m\psi\pd_\n\psi- V(\psi)\right)
\ee
(where $\psi$ generically stands for any matter field), so that the matter potential does not couple to gravitation at the lagrangian level.
Actually things are not so simple though, and there is an interaction forced upon us by the Bianchi identity.
A novel aspect is that the unimodular condition breaks full diffeomorphism invariance $\emph{Diff}(M)$ (where $M$ is the space-time manifold) to a subgroup $\emph{TDiff}(M)$ consisting on those diffeomorphisms $x\rightarrow y$ that have unit jacobian, that is
\be
\text{det}\left({\pd y^\a\over \pd x^\m}\right)=1
\ee
(at the linear level we shall denote these groups by $\emph{LDiff}(M)$ and $\emph{LTDiff}(M)$ respectively). Those constitute the subgroup that leaves invariant the determinant of the metric.
\par
Incidentally, van der Bij,~van Dam and ~Ng \cite{vanderBij} showed a long time ago that $\emph{TDiff}$ is enough to make gauge artifacts of the three excess gauge polarizations when going to the massless limit in a spin two flat space theory (there are five polarizations in the massive case and only two in the massless limit). This intuitively means that we only need three arbitrary gauge parameters, which is exactly what we have in $\emph{LTDiff}(M)$ as will be seen in a moment.
\par

It is possible (and technically convenient) to formulate the theory in such a way that it has an added Weyl invariance by writing
\be
\tilde{g}_{\m\n}\equiv \left|g\right|^{-{1\over n}}~g_{\m\n}
\ee
The reason is that then the variations $\delta g_{\a\b}$ are unconstrained, whereas the variations of the unimodular metric have got to be traceless
\be
\tilde{g}^{\a\b}\delta \tilde{g}_{\a\b}=0
\ee

The plentiful complications brought up by the presence of Weyl symmetry are nevertheless not as nasty as would be the ones stemming from trying to perform the functional integral over unimodular metrics
\be
{\cal D}\tilde{g}_{\m\n }
\ee
which is an integral over constrained functional variables. As a matter of fact, there are maybe other ways of defining the path integral over unimodular metrics; it is then a nontrivial question whether those different path integrals yield physically  equivalent results.
\par
\par 
Let us begin by recalling an analysis of spin two theories in flat space.
In reference \cite{AlvarezBGV} the most general action principle built out of dimension four operators for a spin two field $h_{\m\n}$ was considered, namely
\be
L\equiv \sum_{i=1}^4 C_i ~{\cal O}^{(i)}
\ee
\bea
&&{\cal O}^{(1)}\equiv {1\over 4}\pd_\m h_{\r\s}\pd^\m h^{\r\s}\nonumber\\
&&{\cal O}^{(2)}\equiv -{1\over 2}\pd^\r h_{\r\s}\pd_\m h^{\m\s}\nonumber\\
&&{\cal O}^{(3)}\equiv {1\over 2}\pd_\m h\pd_\l h^{\m\l}\nonumber\\
&&{\cal O}^{(4)}\equiv -{1\over 4}\pd_\m h\pd^\m h
\eea
where all indices are raised and lowered with the flat space metric $\eta_{\m\n}$, and $h\equiv \eta^{\m\n} h_{\m\n}$.
Also $C_1= 1$ fixes the global normalization.
The result of the work in question was that $\emph{LTDiff}$ invariance forces
\be
C_2=1
\ee
where the linearized $\emph{LTDiff}$ invariance is just
\be
\delta h_{\m\n}=\pd_\m \xi_\n+\pd_\n\xi_\m
\ee
with
\be
\pd_\m\xi^\m=0
\ee
The most important result was however the following. Amongst all the $\emph{TDiff}$ invariant theories obtained for arbitrary values of $C_3$ and $C_4$ there are only two that propagate spin two only, without any admixture of spin zero. Those are, first 
\be
C_3=C_4=1
\ee
which has an enhanced symmetry under linearized diffeomorphisms (without the transversality restriction). This is the Fierz-Pauli theory.
\par
The other one corresponds to
\bea
&&C_3={2\over n}\nonumber\\
&&C_4={n+2\over n^2}
\eea
This second theory is actually a truncation of the Fierz-Pauli one obtained by
\be
h_{\m\n}\rightarrow h_{\m\n}-{1\over n} h \eta_{\m\n}
\ee
(which is not a field redefinition, because it is not invertible). This theory was dubbed $\emph{WTDiff}$ and is actually the linear limit of Unimodular Gravity. Let us now turn our attention to the full nonlinear theory. We shall follow the idea of defining the theory starting from General Relativity by a non-invertible field redefinition (namely $g_{\m\n}\rightarrow |g|^{-{1\over n}}g_{\m\n}$), generalizing the way the flat space $\emph{WTDiff}$ theory is defined starting from the Fierz-Pauli theory.

\par
The truncation of General Relativity \footnote{Our conventions are the Landau-Lifshitz spacelike ones. The Minkowski spacetime metric is 
\be
ds^2\equiv \eta_{\m\n} dx^\m dx^\n= dt^2-\sum_{i=1}^{n-1}~\left(dx^i\right)^2 
\ee
Planck's mass is defined as
\be
M_P^{n-2} \equiv {1\over 2\kappa^2}\equiv{c^3\over 16\pi G}
\ee

We will work until the end in arbitrary spacetime dimension $n$. Counterterms however will be computed in the physical dimension n=4.} to unimodular metrics is simply
\bea\label{UG}
&&S_{UG}\equiv - M_P^{n-2}\int d^n x \left(R[\hat{g}]+L_{\text{matt}} [\psi_i,\hat{g}]\right)= \nonumber\\
&&=-M_P^{n-2}\int d^n x ~ |g|^{1\over n}~\left(R+{(n-1)(n-2)\over 4 n^2}{\nabla_\m g\nabla^\m g\over g^2}+ L_{\text{matt}}[\psi_i,|g|^{-{1\over n}}~g_{\m\n}]\right)
\eea
In terms of an unconstrained metric, the equations of motion (EM) are given \cite{AlvarezBGV} by the manifestly traceless expression

\bea
&&R_{\m\n}-\frac{1}{n}R g_{\m\n}-{(n-2)(2n-1)\over 4 n^2} \left({\nabla_\m g\nabla_\n g \over g^2}-{1\over n} {(\nabla g)^2\over g^2} g_{\m\n}\right)+{n-2\over 2n} \left({\nabla_\m \nabla_\n g \over g}-{1\over n} 
{\nabla^2 g\over g} g_{\m\n}\right)=\nonumber\\
&&=M_P^{2-n}\left(T_{\m\n}-\frac{1}{n}T g_{\m\n}\right)
\eea
When $|g|=1$ they are quite similar to the ones posited in 1919 by Einstein for obscure reasons \cite{Einstein,Ellis} related to Mie's theory. These were essentially the tracefree piece of the classic Einstein equations. Einstein himself erroneously thought they were inconsistent \footnote{We are grateful to GFR. Ellis for clarifying this point for us.}.

The Bianchi identities bring the trace back into the game, albeit in a slightly different form. Starting from the second Bianchi identity
\be
\nabla_\m R^{\m\n}={1\over 2}\nabla^\n R
\ee
and taking the value of the covariant derivative of the Ricci tensor from the unimodular equations of motion leads to

\begin{align}
\frac{n-2}{2n}\D_{\m}R=-\frac{\kappa^2}{n}\D_{\m}T
\end{align}
We have assumed that the theory is free from gravitational anomalies, in that the energy momentum tensor of the matter part is covariantly conserved.
The preceding equation implies
\begin{align}
\frac{n-2}{2n}R+\frac{\kappa^2}{n}T=-C\label{Ceq}
\end{align}

Where $C$ is a constant on which more later. Substituting the ensuing value for the trace of the energy-momentum tensor back in the unimodular equations leads to
\begin{align}
R_{\m\n}-\frac{1}{2}R g_{\m\n}-C g_{\m\n}=T_{\m\n}
\end{align}


In order to elucidate the physical meaning of the constant $C$, let us analyse the equations of motion for a Friedmann metric with flat spatial sections with matter content being a scalar field with a constant potential, both in Unimodular Gravity and General Relativity.

 The EM for the scalar field are the same no matter what the theory is,
 \be \Box \phi=0\label{scalar}\ee

We found it convenient to work in the unimodular gauge in which the metric reads

\be ds^2=a(t)^{-\tfrac{3}{2}}dt^2-a(t)^{\tfrac{1}{2}}dx^2\ee

 Now the UG EM reduce just to
 \be  R_{\m\n}-\dfrac{1}{n}Rg_{\m\n}=M_P^{2-n} \left(T_{\m\n}-\dfrac{1}{n}Tg_{\m\n}\right)\label{UGEM}  \ee
The canonical energy-momentum tensor \footnote{It is worth to remark that in Unimodular Gravity the canonical energy momentum tensor
\be T_{\m\n}^\text{can}=\dfrac{\partial L}{\partial \phi_\m}\phi_\n-L g_{\m\n}\label{tcan}\ee
fails to coincide with the variation of the action
\be T_{\m\n}^\text{can}\neq \dfrac{2}{\sqrt{g}}\dfrac{\delta S}{\delta g_{\m\n}}\ee

The reason is obviously that  the particular piece $L g_{\m\n}$ in \eqref{tcan} comes from the variation of $\sqrt{g}$ in the action, which is not present in Unimodular Gravity.}
 is given by
\bea
&&T_{00}=\dfrac{\dot{\phi}^2}{2}+V_0a^{-3/2}\\
&&T_{ij}=\left(a^2\dfrac{\dot{\phi}^2}{2}-V_0a^{1/2}\right)\d_{ij}\\
&&T=\dfrac{2-n}{n}a^{3/2}\dot{\phi}^2+nV_0\eea

The scalar field $\phi$  is related to the scale factor
\be \dot{\phi}=\dot{\phi}_0 a_0^{3/2}a^{-3/2}\equiv K a^{-3/2}\ee

The UG EM read

\be\begin{array}{rr}\dfrac{3\dot{a}^2}{32a^2}-\dfrac{3\ddot{a}}{8a}&=\dfrac{3}{4M_P^{n-2}}\dot{\phi}^2\\
\dfrac{\dot{a}^2}{32}-\dfrac{a\ddot{a}}{8}&=\dfrac{1}{4M_P^{n-2}}\dot{\phi}^2\end{array}\ee
Plugging the first integral for the scalar field, they reduce to
\be
\dot{a}^2-4a\ddot{a}=\dfrac{\tilde{C}}{a}
\ee
where $\tilde{C}=8\dfrac{K^2}{M_P^{n-2}}$.\par

This equation has only one real solution which is

\be
a[t]^\text{UG}=\left(\dfrac{16\tilde{C}-27t^2C_1^2-54tC_1^2C_2-27C_1^2C_2^2}{48 C_1}\right)^{2/3}
\ee

The solution  depends on two integration constants (being determined by a second order differential equation) and the value of $V_0$ is inmaterial for them.\par

Concerning $C$ in \eqref{Ceq} this is just

\be C=\dfrac{3(n-2)}{16n}\dfrac{\dot{a}^2+4a\ddot{a}}{\sqrt{a}}+\dfrac{1}{n}\left(K^2a^{-3/2}+4V_0\right)=\dfrac{3(3n-4)}{16n}C_1
\ee

The GR EM read in turn

\be\begin{array}{ll}
        &       3\dfrac{\dot{a}^2}{16a^2}=\dfrac{1}{M_P^{n-2}}\left(\dfrac{\dot{\phi}^2}{2}+V_0a^{-3/2}\right)\\
        &-\dfrac{1}{16}(\dot{a}^2+8a\ddot{a})=\dfrac{1}{M_P^{n-2}}\left(a^2\dfrac{\dot{\phi}^2}{2}-V_0a^{1/2}\right)\end{array}\ee

It is plain that there is a linear combination of both equations which yields the UG EM,
\be
\dot{a}^2-4a\ddot{a}=\dfrac{\bar{C}}{a}
\ee
 but this is not the whole story. Going back and substituting in the full EM one of the arbitary constants is actually related to the zero mode of the potential energy

\be C_1=\pm \dfrac{16V_0}{3M_P^{n-2}}\ee

Summarizing, in GR there is just one arbitray integration constant; the other one turns out to be (proportional to) the cosmological constant. However, in UG there are two arbitrary integration constants.\par

Although $\Lambda$ fullfills the same equation than C, in this case $C_1$ is not an arbitrary integration constant but proportional to $V_0$ then in the case of GR the putative constant is precisely

\be \Lambda=\pm\dfrac{(3n-4)}{n}\dfrac{V_0}{M_P^{n-2}}
\ee


 The aim of the present work is to examine whether there are quantum corrections to this physical setup. In case they were present (they are not) the importance of the classical result would not be great.
\par
 It is worth remarking that a recent reanalysis of the classical Deser \cite{Deser} argument for the nonlinear completion of the linear Fierz-Pauli theory in \cite{Barcelo} gets the result that both Unimodular Gravity and General Relativity (and only these two) are the allowed possibilities.

The purpose of the present paper is to compute the path integral of Unimodular Gravity
\begin{align}
{\cal Z}=\int {\cal D}g_{\m\n}\;e^{-S_{UG}}
\end{align}
with $S_{UG}$ as given in \eqref{UG}. This is done up to one loop order in the background field formalism, where we expand the classical action around a background metric $\tilde{g}_{\m\n}$ and quantize one-loop fluctuations around this. This will allow us to make sense of a gauge fixing which takes care of the transversality condition imposed by \emph{TDiff} at the cost of introducing new Batalin-Vilkovisky \cite{Batalin} fields which may be interpreted as a combination of Faddeev-Poppov and Nielsen-Kallosh \cite{Siegel} ghosts. Some previous work in Unimodular Gravity can be found in \cite{Eichhorn,Smolin}. The closest in spirit to the present work is \cite{AlvarezFV} where Unimodular Gravity is studied as a gauge fixed theory of a full $\emph{Diff}$ invariant extension using compensator fields.

\section{The backgroung field method and the BRST quantization of Unimodular Gravity}\label{quantum}
In order to compute quantum corrections to Unimodular Gravity, we are going to rely on two main techniques of common use in the path integral approach to gauge theories and, in particular, to quantum gravity: the background field method in conjunction with the BRST formalism and the Schwinger-De Witt technique. Using a combination of both, we will be able to compute the quantum effective action of the theory for arbitrary background solutions of the classical fields. We should like to stress that the application of BRST formalism to the case at hand is quite involved 
 since, as we shall see below, the theory has first-stage reducible gauge transformations in the language of Ref.~\cite{Batalin}. Further, the choice of gauge-fixing terms has to be made with care; otherwise one ends up dealing with very complicated differential operators, which puts the feasibility of the one-loop computation in jeopardy.
 
To quantize the classical Unimodular Gravity theory defined by the action in \eqref{UG} within the background field formalism, one splits the
metric $g_{\mu\nu}$ into two parts: one contains the background metric $\bar{g}_{\mu\nu}$
 and the other the quantum fluctuations $h_{\mu\nu}$. We shall find it advantageous
to use the following splitting
\begin{align}\label{newsplit}
g_{\mu\nu}= \bar{g}_{\mu\nu}+ |\bar{g}|^{\frac{1}{n}}\, h_{\mu\nu}
\end{align}
rather than usual splitting $g_{\mu\nu}= \bar{g}_{\mu\nu}+ h_{\mu\nu}$. Notice that we can convert the splitting in \eqref{newsplit} into the usual splitting by performing a Weyl transformation of the quantum field. This is supported by the claim in \cite{Alvarezhv} that there is no conformal anomaly in Unimodular Gravity.

Let us also express the background metric $\bar{g}_{\mu\nu}$ in terms of a metric $\tilde{g}_{\mu\nu}$ such that $|\tilde{g}|=1$ as follows
\begin{align}\label{bg_transf}
\bar{g}_{\mu\nu}= |\bar{g}|^{\frac{1}{n}}\,\tilde{g}_{\mu\nu}
\end{align}
Then, we have the following equality which stems from the Weyl invariance of the classical action
\begin{align}\label{backaction}
S_{UG}[\,g_{\mu\nu}= \bar{g}_{\mu\nu}+ |\bar{g}|^{\frac{1}{n}}\, h_{\mu\nu}]=S_{UG}[\,g_{\mu\nu}= \tilde{g}_{\mu\nu}+ h_{\mu\nu}]
\end{align}
where $S_{UG}[\,g_{\mu\nu}]$ is given in \eqref{UG}.
\par

Let us warn the reader that from now on the covariant derivative will be defined with respect to the metric $\tilde{g}_{\mu\nu}$ and that, unless explicitly said, we are dropping the tilde over background quantities in order to get cleaner formulas. Conceptually however it is important to keep in mind that there are two different classical metrics so far; $\bar{g}_{\m\n}$ representing an {\em arbitrary} background metric and $\tilde{g}_{\m\n}$ a unimodular ($\tilde{g}=-1$) background metric.
\par

Thus, from now we will write
\begin{align}
\tilde{g}_{\m\n}=g_{\m\n}
\end{align}

To quantize the theory defined by $S_{UG}$ in \eqref{backaction}, one has to identify first the gauge symmetries of it and then fix them. It can be seen that $S_{UG}$ is invariant
 under the following --written in BRST form-- gauge transformations
 \begin{align}\label{tdiff}
&s_{D}g_{\m\n}=s_{W}g_{\m\n}=0\nonumber\\
\nonumber &s_{D}h_{\m\n}=\D_{\m}c^{T}_{\n}+\D_{\n}c^{T}_{\m}+c^{T\r}\D_{\r}h_{\m\n}+\D_{\m}c^{T\r}  h_{\r\n} + \D_{\n}c^{T\r}  h_{\r\m}\\
&s_{W}h_{\m\n}=2c\left(g_{\m\n}+h_{\m\n}\right)
\end{align}
where $c$ and $c^{T\m}$ are the anticommuting ghost fields for Weyl invariance and transverse diffeomorphisms, respectively. In this language, the transverse condition is satisfied by imposing $\D_{\m}c^{T\m}=0$ on the ghost field. The superscript $T$ thus means that the vector satisfies this condition. The gauge fixing procedure of these gauge symmetries
will be discussed next.
\par
The partition function of the theory is now
\begin{align}
{\cal Z}[J]=\int {\cal D}h_{\m\n}\;e^{-S_{2}-\int d^{n}x\;J^{\m\n}h_{\m\n}}
\end{align}
and when the sources vanish this defines the quantum effective action to be
\begin{align}
W=-\frac{1}{2}\log\left(\det{\cal D}\right)
\end{align}
where ${\cal D}$ is the operator driving the one-loop quantum fluctuations, defined by the quadratic term in the expansion of the action around the background metric
\begin{align}
S_{2}=\int d^{n}x\;{\cal L}_2=\int d^{n}x\;h^{\m\n}{\cal D}_{\m\n\r\s}h^{\r\s}
\end{align}
It is useful to write down the expression as it would stand {\em before} the background metric is assumed to be unimodular
\begin{align}
\nonumber &{\cal L}_{2}=\frac{1}{4}h^{\m\n}\overline{\Box}h_{\m\n}-\frac{n+2}{4 n^{2}}h \overline{\Box}h  +\frac{1}{2} \left(\bD_{\m}h^{\m\a}\right)\left(\bD_{\n}h^{\n}_{\a}\right)-\frac{8-6n +n^{2}}{8 n^{2}}\left(\bD^{\a}\log \bg\right)\;  h^{\m\n}\bD_{\a}h_{\m\n}-\\
\nonumber& -\frac{1}{n}\left(\bD_{\m}h\right)\left(\bD_{\n}h^{\m\n}\right)+\frac{2-n}{2n}\left(\bD_{\a}\log \bg\right)\left(\frac{1}{2} h^{\b\lambda}\bD_{\lambda}h^{\a}_{\b}+\frac{3}{2}h^{\a\b}\bD_{\lambda}h^{\lambda}_{\b}-\frac{1}{n}h\bD_{\lambda}h^{\a\lambda}\right)+\\
\nonumber&+\frac{(n-2)^2}{8 n^3} \left(\bD^{\a}\log \bg\right)\; h \bD_{\a}h+\frac{n-2}{2 n^2}\left( \bD_{\b}\log \bg\right)\; h^{\a\b}\bD_{\a}h+\frac{1}{2}h^{\a\b}h^{\m}_{\b}\bR_{\m\a}-\frac{1}{n}h h^{\m\n}\bR_{\m\n}+\\
\nonumber &+(n^2  -3n  +2)\left( \frac{1}{8 n^4} h^2  (\bD\log \bg)^2  -\frac{1}{8 n^3}h_{\m\n}h^{\m\n}   (\bD\log \bg)^2  +\frac{1}{4 n^2} h_{\a}^{\lambda}h_{\b\lambda} \left(\bD^{\a}\log \bg\right)   \left(\bD^{\b}\log \bg\right)   -\right. \\
&\left. -\frac{1}{4  n^3}h h_{\a\b}\left(\bD^{\a}\log \bg\right)  \; \left(\bD^{\b}\log \bg\right) \right)+\frac{1}{2}h^{\m\n}h^{\a\b}\bR_{\m\a\n\b}-\frac{1}{2n}h^{\m\n}h_{\m\n}\bR  +\frac{1}{2 n^{2}}h^{2} \bR
\end{align}

Of course, ${\cal D}$ will contain in principle zero modes coming from the gauge symmetries of the theory translated to the linear level which will make its determinant singular. This is solved by constructing an appropriate gauge fixing term using the BRST quantization method.

Finally, since we are using the splitting \eqref{newsplit}, the action for the one-loop quantum fluctuations simplifies somewhat, since all terms depending on $\D_{\m} g $ now vanish. Thus, we end up with
\begin{align}
 &{\cal L}_{2}=\frac{1}{4}h^{\m\n}\Box h_{\m\n}-\frac{n+2}{4 n^{2}}h \Box h  +\frac{1}{2} \left(\D_{\m}h^{\m\a}\right)\left(\D_{\n}h^{\n}_{\a}\right) -\frac{1}{n}\left(\D_{\m}h\right)\left(\D_{\n}h^{\m\n} \right)+\nonumber\\
&+\frac{1}{2}h^{\a\b}h^{\m}_{\b} R_{\m\a}-\frac{1}{n}h h^{\m\n} R_{\m\n}+\frac{1}{2}h^{\m\n}h^{\a\b} R_{\m\a\n\b}-\frac{1}{2n}h^{\m\n}h_{\m\n} R  +\frac{1}{2 n^{2}}h^{2}  R
\end{align}
 After computing the quantum effective action and owing to the already mentioned fact that no conformal anomaly is present in the theory, one can just undo the transformation \eqref{bg_transf} and recover the expression for arbitrary background metrics by performing a conformal transformation away from the Einstein frame.

\subsection{Fixing the gauge freedom}
To gauge-fix the gauge symmetries in \eqref{tdiff}, we shall use the BRST technique in a similar way as in \cite{AlvarezHVM} and introduce the following nilpotent BRST operator
\begin{align}
s=s_{D}+s_{W}
\end{align}
where $s_{D}$ and $s_{W}$ are defined in \eqref{tdiff}.

The path integral over the ghost fields must be restricted to the subspace of transverse vectors. However, the definition of such a measure $[{\cal D}c^{T\m}]$ over transverse vectors is a notorious problem \cite{Siegel}. The way to come to grips with it chosen in this paper is to parametrize this subspace in terms of unconstrained fields so that we can then integrate over the full space of $c^{\m}$, whose integration measure is well-defined. This we do by introducing an operator $\Theta_{\m\n}$\footnote{One can easily check that $\Theta_{\m\n}$ is indeed an endomorphism in the space spanned by transverse vectors.}
\begin{align}\label{cTmu}
c_{\m}^{T}=\Theta_{\m\n}c^{\n}=\left(g_{\m\n}\Box -\D_{\m}\D_{\n}-R_{\m\n}\right)c^{\n}=\left(Q_{\m\n}-\D_{\m}\D_{\n}\right)c^{\n}
\end{align}
which maps vectors into transverse vectors. In this way, the transversality condition over $c_{\m}^{T}$ translates into a gauge symmetry for $c_{\n}$
\begin{align}\label{firststage}
c_{\n}\rightarrow\D_{\n}f
\end{align}
with $f$ an arbitrary function. Indeed, this transformation takes $c_{\n}$ into a longitudinal vector, so that the $\Theta_{\m\n}$ operator annihilates it. Of course, in order to perform now the functional integration over $c^{\m}$ we must gauge fix this new gauge symmetry by introducing a non-trivial stairway of ghost levels with BRST transformations defined in such a way that the BRST algebra closes
\begin{align}\label{s2=0}
&s_{D}^{2}=s_{W}^{2}\nonumber\\
&\left\{ s_{D},s_{W}\right\}=0
\end{align}
on all the different fields considered.

The systematic way to obtain this field content together with the appropriate BRST transformations is by using the Batalin-Vilkovisky~\cite{Batalin} formalism. However, in our case, things are easy enough as to allow us to guess what the BRST transformations read, once the field
content of the theory is chosen as done in \cite{Batalin} for first-stage reducible and irreducible gauge transformations. Notice that the gauge transformations in \eqref{tdiff} generated by $s_D$, with $c^{}_\mu$ in \eqref{cTmu}, are first-stage reducible due to the gauge symmetry in \eqref{firststage}. However, the gauge symmetries in \eqref{tdiff} generated by $s_W$ are irreducible. We introduce the following set of fields:
\begin{align}
&h^{(0,0)}_{\m\n},\; c^{(1,1)}_{\m},\; b_{\m}^{(1,-1)},\; f_{\m}^{(0,0)},\; \f^{(0,2)},\nonumber\\
&\pi^{(1,-1)},\; \pi'^{(1,1)},\; \bar{c}^{(0,-2)},\; c'^{(0,0)},\nonumber\\
&c^{(1,1)},\; b^{(1,-1)},\;f^{(0,0)}
\end{align}
where $c^{(1,1)}_{\m}$ denotes $c_\mu$, $h^{(0,0)}_{\m\n}$ stands for $h_{\m\n}$ and the superscript $(n,m)$ carries the Grassmann number, $n$, (defined modulo two) and ghost number, $m$. In this language, the BRST operators $s_D$ and $s_W$ enjoy Grassmann number 1 and ghost number 1, each.

Here we have three families --displayed in three different lines-- of fields. The first line includes the physical graviton field together with the usual ghost field content that would be naively necessary in order to gauge fix an unrestricted $\emph{Diff}$ symmetry. In addition, there is a $\f$ field which accounts for the transformation in
\eqref{firststage}. The second line represents the field content introduced to gauge fix the gauge symmetry in \eqref{firststage}, together with the one that will be induced on $b_{\m}^{(1,-1)}$. Finally, the third line is the field content due to Weyl invariance.

\begin{table}
\begin{center}
\begin{tabular}{ c || c | c }
  field & $s_{D}$ & $s_{W}$ \\ \hline \hline
  $g_{\m\n}$ & 0  & 0 \\
  $h_{\m\n}$ & $\D_{\m}c_{\n}^{T}+\D_{\n}c_{\m}^{T}+c^{\rho T}\D_{\r}h_{\m\n}+\D_{\m}c^{\r T}h_{\r\n}+\D_{\n}c^{\r T}h_{\r\m}$ &  $2 c^{(1,1)}\left(g_{\m\n}+h_{\m\n}\right)$\\
 $c^{(1,1)\m}$ &$\left(Q^{-1}\right)^{\m}_{\n}\left(c^{\r T}\D_{\r}c^{T\n}\right)+\D^{\m}\f^{(0,2)}$ & 0\\
 $\phi^{(0,2)}$&0&0\\
 $b_{\m}^{(1,-1)}$&$f_{\m}^{(0,0)}$&0\\
 $f_{\m}^{(0,0)}$&0&0\\
 $\bar{c}^{(0,-2)}$& $\pi^{(1,-1)}$&0\\
 $\pi^{(1,-1)}$&0&0\\
 $c'^{\;(0,0)}$&$\pi'^{\;(1,1)}$&0\\
 $\pi'^{\;(1,1)}$&0&0\\
 $c^{(1,1)}$&$c^{T\rho}\D_{\rho}c^{(1,1)}$&0\\
 $b^{(1,-1)}$&$c^{T\rho}\D_{\rho}b^{(1,-1)}$ &$f^{(0,0)}$\\
 $f^{(0,0)}$&$c^{T\rho}\D_{\rho}f^{(0,0)}$&0\\

\end{tabular}
\caption{BRST transformations of the fields involved in the path integral.}
\label{BRST_TD}
\end{center}
\end{table}

Now, we define the action of $s_D$ and $s_W$ on the fields as shown in Table \ref{BRST_TD}, where the $\left(Q^{-1}\right)^{\m}_{\n}$ denotes the inverse of the operator $Q_{\m\n}=g_{\m\n}\Box - R_{\m\n}$, which exists provided $Det(Q)\neq0$. This is our case since $Q_{\m\n}$ is just a standard Laplacian-type operator acting on vector fields.

With these definitions, it can be readily shown that the equations in \eqref{s2=0} hold. In doing so, it is advisable to show first that
\begin{align}
s_D c^{T\mu} = c^{T\r}\D_{\r}c^{T\n}
\end{align}
if $c^{T\mu}$ is defined as in \eqref{cTmu}. This can be done by using the following results
\begin{align}
\D_\mu( c^{\r T}\D_{\r}c^{T\m})=0,\quad \D_\mu\left[ \left(Q^{-1}\right)^{\m}_{\n}\left(c^{\r T}\D_{\r}c^{T\n}\right)\right]=0
\end{align}

The path integral quantization of the theory is accomplished now by adding to the classical action the gauge-fixing action, $S_{gauge-fixing}$, which is an appropriate BRST-exact term:
\begin{align}
S_{gauge-fixing}=\int d^{n}x\; s \left(X_{TD}+X_{W}\right)
\end{align}
$X_{TD}$ and $X_{W}$ are polynomials of the quantum fields with ghost number -1 and Grassmann number equal to 1 and such that they give rise to free-kinetic terms that are invertible. Since we are only interested
in one-loop computations, we shall further assume that $X_{TD}$ and $X_{W}$ are quadratic in the quantum fields. In the next sections we will construct the terms $X_{TD}$ and $X_W$ and derive the differential operators involved in the path integral whose contribution to the quantum effective action needs to be computed.

\subsection{The $\emph{TDiff}$ sector}
Let us start with the function $X_{TD}$ performing the gauge fixing of the $\emph{TDiff}$ symmetry. With the field content introduced above and with the BRST transformations as given in table \ref{BRST_TD}, one has the following general quadratic polynomial in the quantum fields associated to the gauge-fixing of the $\emph{TDiff}$ symmetry

\begin{align}
\nonumber X_{TD}&=b_{\m}^{(1,-1)}\left[F^{\m}+\rho_{1}f^{\m(0,0)}\right]+\bar{c}^{(0,-2)}\left[F^{\m}_{2}c_{\m}+\rho_{2}\pi'^{'\;(1,1)}\right]+\\
&+c'^{\;(0,0)}\left[F_{1}^{\m}b_{\m}^{(1,-1)} +\rho_{3}\pi^{(1,-1)}\right]
\end{align}
where $F_{\m}$ is a function containing the graviton field that can be identified with the usual gauge fixing condition in the Faddeev-Poppov technique and $F_{1}^{\m}$, $F_{2}^{\m}$ and the three $\rho_{i}$ can be freely chosen. This is enough to fix the $\emph{TDiff}$ symmetry with the minimal possible content of fields.

After applying the $s$ operator, this gives a term in the action
\begin{align}
\nonumber \int d^{n}x\;  s X_{TD}&=\int d^{n}x\; f^{(0,0)}_{\m}\left(F^{\m}+\rho_{1}f^{\m(0,0)}\right)-b_{\m}^{(1,-1)}sF^{\m}+\\
\nonumber &+\pi^{(1,-1)}\left(F_{2}^{\m}c_{\m}^{(1,1)}+\rho_{2}\pi'^{\;(1,1)}\right)+\bar{c}^{(0,-2)}F_{2}^{\m}\D_{\m}\f^{(0,2)}+\\
&+\pi'^{\;(1,1)}\left(F_{1\m}b^{\m (1,-1)}+\rho_{3}\pi^{(1,-1)}\right)+c'^{\;(0,0)}F_{1}^{\m}f_{\m}^{(0,0)}
\end{align}
where we have already taken into account that in the expansion \eqref{newsplit} the metric is unimodular.

Now, there are some simplifications that can be done. First, let us take the terms containing $f_{\m}^{(0,0)}$
\begin{align}
f_{\m}^{(0,0)}\left(F^{\m}+\rho_{1}f^{\m (0,0)}\right)+f^{(0,0)}_{\m}\bar{F}_{1}^{\m}c'^{\;(0,0)}
\end{align}
where we have introduced $\bar{F}_{1}^{\m}$ using integration by parts as
\begin{align}
\int d^{n}x\; a F_{1}^{\m}b=\int d^{n}x\; b\bar{F}_{1}^{\m}a
\end{align}

These can be rewritten completing the square as
\begin{align}
\rho_{1}\left(f_{\m}^{(0,0)}+\frac{1}{2 \rho_{1}}(F_{\m}+\bar{F}_{1 \m}c'^{\;(0,0)})\right)^{2}-\frac{1}{4\rho_{1}}(F_{\m}+\bar{F}_{1\m}c'^{\;(0,0)})^{2}
\end{align}
and shifting the variable $f_{\m}^{(0,0)}$ the first term does not contribute to the effective action and we are left with the gauge fixing action
\begin{align}
S_{hc'}=-\frac{1}{4\rho_{1}}\int d^{n}x\;(F_{\m}+\bar{F}_{1\m}c'^{\;(0,0)})^{2}
\end{align}
where $\rho_{1}$ has been chosen to be a constant. This would be the outcome of a standard Faddeev-Poppov procedure.

Now let us focus into the terms containing the fermionic $\pi$ fields. Those read
\begin{align}
\nonumber &\pi^{(1,-1)}\left(F_{2}^{\m}c_{\m}^{(1,1)}+\rho_{2}\pi'^{\; (1,1)}\right)+\pi'^{\;(1,1)}\left(F^{\m}_{1}b_{\m}^{(1,-1)}+\rho_{3}\pi^{(1,-1)}\right)=\\
\nonumber &=\left(\pi^{(1,-1)}-F_{1}^{\m}b_{\m}^{(1,-1)}(\rho_{2}-\rho_{3})^{-1}\right)(\rho_{2}-\rho_{3})\left(\pi'^{\;(1,1)}+(\rho_{2}-\rho_{3})^{-1}F_{2}^{\m}c_{\m}^{(1,1)}\right)+\\
&+F_{1}^{\m}b_{\m}^{(1,-1)}(\rho_{2}-\rho_{3})^{-1}F_{2}^{\m}c_{\m}^{(1,1)}
\end{align}
and, again, by shifting the $\pi$ fields we are left with a gauge fixing term plus an extra path integral depending on how we choose the operators $\rho_{2}$ and $\rho_{3}$
\begin{align}
S_{\pi}+S_{gf}^{bc}=\int d^{n}x\;\left(\pi^{(1,-1)}(\rho_{2}-\rho_{3})\pi'^{\;(1,1)}+F_{1}^{\m}b_{\m}^{(1,-1)}(\rho_{2}-\rho_{3})^{-1}F_{2}^{\m}c^{(1,1)}_{\m}\right)
\end{align}

So that the BRST action for the $\emph{TDiff}$ sector is further simplified to
\begin{align}
\nonumber\int d^{n}x\;  s X_{TD}=&\int d^{n}x\;\left(-b_{\m}^{(1,-1)}sF^{\m}+\bar{c}^{(0,-2)}F_{2}^{\m}\D_{\m}\f^{(0,2)}+\pi^{(1,-1)}(\rho_{2}-\rho_{3})\pi'^{\;(1,1)}+\right.\\
&\left.+F_{1}^{\m}b_{\m}^{(1,-1)}(\rho_{2}-\rho_{3})^{-1}F_{2}^{\m}c^{(1,1)}_{\m}-\frac{1}{4\rho_{1}}(F_{\m}+\bar{F}_{1\m}c'^{\;(0,0)})^{2}\right)
\end{align}

As a next step, the function $F_{\m}$ is chosen with two requirements in mind. First, that the term $F_{\m}F^{\m}$ is able to cancel the non-diagonal pieces of the operators in the original lagrangian for the graviton fluctuations and also that it is Weyl invariant so both gauge fixing sectors decouple and their ghost fields do not interact. With these two requirements, the choice is almost unique
\begin{align}
F_{\m}=\D^{\n}h_{\m\n}-\frac{1}{n}\D_{\m}h
\end{align}
and its variation under a transverse diffeomorphism is the equivalent to the application of the $s$ operator
\begin{align}
s F_{\m}=\Box c_{\m}^{T}+\D^{\n}\D_{\m}c_{\n}^{T}=\Box c_{\m}^{T}+R_{\m}^{\n}c_{\n}^{T}
\end{align}
where in the second step we have used Ricci identity $[\D_{\n},\D_{\m}]c^{\n}=R_{\m\n}c^{\n}$ and the fact that, since we are performing a transverse diffeomorphism, $c_{\m}^{T}$ satisfies $\D^{\m}c_{\m}^{T}=0$.

Now, we have to rewrite $c_{\m}^{T}$ in terms of an unconstrained field as explained before. We do this by introducing the operator $\Theta_{\m\n}$.
\begin{align}
\nonumber s F_{\m}&=\left(g_{\m}^{\a}\Box+R_{\m}^{\a}\right)\left(g_{\a\n}\Box -\D_{\a}\D_{\n}-R_{\a\n}\right)c^{\n(1,1)}=\\
\nonumber &=\Box^{2}c^{(1,1)}_{\m}-\D_{\m}\Box \D_{\n}c^{\n(1,1)}-2R_{\m\r}\D^{\r}\D_{\n}c^{\n(1,1)}-\Box R_{\m\r}c^{\r(1,1)}-\\
&-2\D_{\s}R_{\m\r}\D^{\s}c^{\r(1,1)}-R_{\m\r}R^{\r\n}c_{\n}^{(1,1)}
\end{align}

The action for $b_{\m}^{(1,-1)}$ and $c^{(1,1)}_{\m}$ is
\begin{align}
\nonumber S_{bc}=-\int d^{n}x\;b^{\m\,(1,-1)}&\left(\Box^{2}c^{(1,1)}_{\m}-\D_{\m}\Box \D_{\n}c^{\n(1,1)}-2R_{\m\r}\D^{\r}\D_{\n}c^{\n(1,1)}-\Box R_{\m\r}c^{\r(1,1)}-\right.\\
&\left. -2\D_{\s}R_{\m\r}\D^{\s}c^{\r(1,1)}-R_{\m\r}R^{\r\n}c_{\n}^{(1,1)}\right)
\end{align}

The non-diagonal term with four derivatives can be canceled by an appropriate choice of the functions $F_{1}^{\m}$, $F_{2}^{\m}$, $\rho_{2}$ and $\rho_{3}$. We choose them to be
\begin{align}
&F_{1}^{\m}b^{(1,-1)}_{\m}=-\D^{\a}b^{(1,-1)}_{\a}\nonumber\\
&F_{2}^{\m}c^{(1,1)}_{\m}=\D^{\m}c^{(1,1)}_{\m}\nonumber\\
&(\rho_{2}-\rho_{3})^{-1}=-\Box
\end{align}
Then
\begin{align}
&F_{1}^{\m}b_{\m}^{(1,-1)}(\rho_{2}-\rho_{3})^{-1}F_{2}^{\m}c_{\m}^{(1,1)}=\left(\D^{\n}b^{(1,-1)}_{\n}\right)\Box \D^{\m}c^{(1,1)}_{\m}=-b^{(1,-1)}_{\n}\D^{\n}\Box \D^{\m}c^{(1,1)}_{\m}
\end{align}
where in the second step we have performed an integration by parts keeping in mind that we are always under an integral sign. The final action term for $b_{\m}^{(1,-1)}$ and $c^{(1,1)}_{\m}$ is then
\begin{align}\label{Sbc}
\nonumber S_{bc}+S_{gf}^{bc}=\int d^{n}x\; b^{\m\,(1,-1)}&\left(\Box^{2}c^{(1,1)}_{\m}-2R_{\m\r}\D^{\r}\D_{\n}c^{\n(1,1)}-\Box R_{\m\r}c^{\r(1,1)}-\right.\\
&\left. -2\D_{\s}R_{\m\r}\D^{\s}c^{\r(1,1)}-R_{\m\r}R^{\r\n}c_{\n}^{(1,1)}\right)
\end{align}

And with this choice of $(\r_{2}-\r_{3})$, the integration over the $\pi$ fields is given by
\begin{equation}\label{Spi}
S_{\pi}=\int d^{n}x\;\pi^{(1,-1)}\Box^{-1}\pi'^{\;(1,1)}
\end{equation}
\par
The operator involving $c'^{\; (0,0)}$, and induced by this choice of fixing functions is
\begin{align}
S_{hc'}&=-\int d^{n}x\;\frac{1}{4\rho_{1}}\left[\bar{F}_{1}^{\m}c'^{(0,0)}\bar{F}_{1\m}c'^{(0,0)}+2F_{\m}\bar{F}_{1}^{\m}c'^{\;(0,0)}+F_\m F^\m\right]=\nonumber\\
&=-\int d^{n}x\;\frac{1}{4\rho_{1}}\left[\D_{\m}c'^{\;(0,0)}\D^{\m}c'^{\;(0,0)}+2 F_{\m}\D^{\m}c'^{\;(0,0)}+F_\m F^\m\right]
\end{align}
which mixes with the operator of the graviton fluctuation due to the term containing $F_{\m}$ and $c'^{\;(0,0)}$.

Finally, the operator for $\bar{c}^{(0,-2)}$ and $\f^{(0,2)}$ is
\begin{align}\label{Scphi}
S_{\bar c\f}=\int d^{n}x\; \bar{c}^{(0,-2)} \Box \f^{(0,2)}
\end{align}

Summarizing, the BRST exact action for the TDIff symmetry is reduced to
\begin{align}
S_{BRST}^{\emph{TDiff}}=&\int d^{n}x\;b^{\m}\left(\Box^{2}c^{(1,1)}_{\m}-2R_{\m\r}\D^{\r}\D_{\n}c^{\n(1,1)}-\Box R_{\m\r}c^{\r(1,1)}-\nonumber\right.\\
&\left. -2\D_{\s}R_{\m\r}\D^{\s}c^{\r(1,1)}-R_{\m\r}R^{\r\n}c_{\n}^{(1,1)}\right)+\bar{c}^{(0,-2)} \Box \f^{(0,2)}+\pi^{(1,-1)}\Box^{-1}\pi'^{\;(1,1)}-\nonumber\\
&-\frac{1}{4\rho_{1}}\left(F_{\m}F^{\m}+\D_{\m}c'^{\;(0,0)}\D^{\m}c'^{\;(0,0)}+2 F_{\m}\D^{\m}c'^{\;(0,0)}\right)=\nonumber\\
&=S_{bc}+S_{gf}^{bc}+S_{\bar{c}\f}+S_{\pi}+S_{h c'}
\end{align}

The contribution of all these pieces to the quantum effective action will be computed in section \ref{effective-action}.

\subsection{The Weyl sector}
Now we turn our attention to the second part of the gauge fixing sector, corresponding to the Weyl invariance of the theory. We choose the function $X_{W}$ to be
\begin{align}
X_{W}=\D_{\m}b^{(1,-1)}\D^{\m}\left(f^{(0,0)}-\a\; g(h)\right)
\end{align}
with $g(h)$ being some function of the trace of the graviton fluctuation only, to ensure that it is invariant under a $\emph{TDiff}$ transformation. The parameter $\a$ we mean to keep arbitrary all along the computation. The on shell effective action should be independent of $\a$ (because it appears in a BRST exact piece), and this will be used as a nice partial check of our results.

After the application of $s$, the BRST exact action is
\begin{align}
S_{BRST}^{Weyl}=\int d^{n}x\;\left[\D_{\m}f^{(0,0)}\D^{\m}\left(f^{(0,0)}-\alpha\; g(h)\right)-\alpha\D_{\m}b^{(1,-1)}\D^{\m} \left(sg(h)\right)\right]
\end{align}

And we choose $g(h)$ to be the simplest choice
\begin{align}
g(h)=h
\end{align}

The BRST term piece is then
\begin{align}\label{SW}
\nonumber S_{Weyl}&=\int d^{n}x\;\D_{\m}f^{(0,0)}\D^{\m}\left(f^{(0,0)}-\alpha \; h\right)-2n\alpha\D_{\m}b^{(1,-1)}\D^{\m}c^{(1,1)}=\\
&=\int d^{n}x\;\left(-f^{(0,0)}\Box f^{(0,0)} +\frac{\alpha}{2}f^{(0,0)}\Box h+\frac{\alpha}{2}h\Box f^{(0,0)}\right)+2n\a\;b^{(1,-1)}\Box c^{(1,1)}=S_{W}+S_{hf}
\end{align}

This gives two contributions to the one-loop effective action. The first part needs to be added to the original action of Unimodular Gravity. The second piece is the corresponding ghost action.

\section{The one-loop effective action of Unimodular Gravity}\label{effective-action}
Once the gauge freedom is fixed completely, the computation of the one-loop counterterm of Unimodular Gravity is reduced to a computation of a set of determinants. By collecting all the terms defined in the previous sections, the pole part of the one-loop effective action will be given, as explained in appendix \ref{heat_kernel}, by
\begin{align}
W_{\infty}=W_{\infty}^{UG}+W_{\infty}^{bc}+W_{\infty}^{\pi}+W_{\infty}^{\bar{c}\f}+W_{\infty}^{W}
\end{align}
where each $W_{\infty}^{i}$ refers to the contribution to the pole given by the action labeled as $S_{i}$ in the previous sections, with the only exception of $W_{\infty}^{UG}$ which is given by
\begin{align}\label{Sug}
S_{UG}=S_{2}+S_{hc'}+S_{hf}
\end{align}

Each of this action terms have the general structure
\begin{align}\label{form_operator}
S=\int d^{n}x\; \Psi^{A}F_{AB}\Psi^{B}
\end{align}
where $\Psi^{A}$ will be a vector containing different fields and $F_{AB}$ a differential operator action over the fields. For instance, if we take $S_{W}$ we identify
\begin{align}
\Psi^{A}=\binom{b}{c}
\end{align}
and the operator to be 
\begin{align}
F_{AB}=\begin{pmatrix}
0 & 1\\ 
1 & 0
\end{pmatrix}\times n\alpha \Box
\end{align}

All but one of the operators involved in our computation are {\em minimal} operators, meaning that their principal symbol is diagonal and they are of the form
\begin{align}
F_{AB}=\gamma_{AB}\Box^{m}+K_{AB}
\end{align}
where $\gamma_{AB}$ is a metric in configuration space (this includes the spacetime metric as well as a metric defined on whatever space in which the indices carried by the fields live) and $K_{AB}$ is a differential operator of order $m-1$ as most. The contribution of an operator of this kind to the quantum effective action is quite standard and their computation was reviewed in \cite{Barvinsky} by using the Schwinger-DeWitt technique. Some details are given in the appendix. Let us explain here the main points of the computation pertaining to the only non-minimal operator, namely the one contained in $S_{UG}$

\begin{align}\label{Sug}
S^{(1)}_{UG}=S_{2}+S_{hc'}+S_{hf}=\int d^{n}x\;{\cal L}
\end{align}
and
\begin{align}
\nonumber {\cal L}&=\frac{1}{4}h^{\m\n}\Box h_{\m\n}-\frac{1}{4 n}h \Box h +\frac{1}{2}h^{\a\b}h^{\m}_{\b}R_{\m\a}+\frac{1}{2}h^{\m\n}h^{\a\b}R_{\m\a\n\b}-\frac{1}{n}h h^{\m\n}R_{\m\n}-\frac{1}{2n}h^{\m\n}h_{\m\n}R  +\\
\nonumber&+\left(-f\Box f +\frac{\alpha}{2}f\Box h+\frac{\alpha}{2}h\Box f\right)-\frac{1}{2}\left(\D_{\m}c'^{\;(0,0)}\D^{\m}c'^{\;(0,0)}+2\left(\D_{\n}h^{\n}_{\m}-\frac{1}{n}\D_{\m}h\right)\D^{\m}c'^{\;(0,0)}\right)+\\
 &+\frac{1}{2 n^{2}}h^{2} R
\end{align}
where $\rho_{1}=\frac{1}{2}$ in order to cancel the non-diagonal parts in the kinetic term for $h_{\m\n}$.

To write it in the form \eqref{form_operator}, we identify
\begin{align}
\Psi^{A}=\begin{pmatrix}
h^{\m\n}\\
f\\
c'\\
\end{pmatrix}
\end{align}
and the differential operator takes the form
\begin{align}
F_{AB}=\gamma_{AB}\Box + J^{\m\n}_{AB}\D_{\m}\D_{\n}+M_{AB}
\end{align}
where the different matrices involved read
\begin{align}
\gamma_{AB}=\begin{pmatrix}
-\frac{1}{4}\left(\frac{1}{4}{\cal K}_{\m\n\r\s}^{\a\b}-{\cal P}_{\m\n\r\s}^{\a\b}\right)g_{\a\b} \quad&\frac{\a}{2}g_{\m\n}& -\frac{1}{8}g_{\m\n}\\ 
\frac{\a}{2} g_{\r\s}  & -1 &0\\
 -\frac{1}{8}g_{\r\s}&0&\frac{1}{2} 
\end{pmatrix}
\end{align}

\begin{align}
J^{\a\b}_{AB}=\begin{pmatrix}
0&0&\frac{1}{4}\left(g^{\a}_{\m}g^{\b}_{\n}+g^{\a}_{\n}g^{\b}_{\m}\right)\\
0&0&0\\
\frac{1}{4}\left(g^{\a}_{\r}g^{\b}_{\s}+g^{\a}_{\s}g^{\b}_{\r}\right)&0&0
\end{pmatrix}
\end{align}

\begin{align}
 &M_{AB}=\begin{pmatrix}
M_{hh}&0&0\\
0&0&0\\
0&0&0
\end{pmatrix}
\end{align}
with
\begin{align}
&M_{hh}=\left(\frac{1}{2}{\cal P}_{\m\n\r\s}^{\a\b}-\frac{1}{4}{\cal K}_{\m\n\r\s}^{\a\b}\right)R_{\a\b}-\frac{1}{8}\left({\cal P}_{\m\n\r\s}^{\a\b}-\frac{1}{4}{\cal K}_{\m\n\r\s}^{\a\b}\right)\gamma_{\a\b}R+\frac{1}{2}R_{(\m\r\n\s)}
\end{align}
The round parenthesis for us mean complete symmetrization in all the enclosed indices unless otherwise stated.
We have introduced the tensors
\begin{align}
{\cal P}_{\m\n\r\s}^{\a\b}&=\frac{1}{4}\left(g_{\m\r}\delta_{\n}^{(\a}\delta_{\s}^{\b)}+g_{\m\s}\delta^{(\a}_{\n}\delta^{\b)}_{\r}+g_{\n\r}\delta^{(\a}_{\m}\delta^{\b)}_{\s}+g_{\n\s}\delta^{(\a}_{\m}\delta^{\b)}_{\r}\right)\\
 {\cal K}_{\m\n\r\s}^{\a\b}&=\frac{1}{2}\left(g_{\m\n}\delta^{(\a}_{\r}\delta^{\b)}_{\s}+g_{\r\s}\delta^{(\a}_{\m}\delta^{\b)}_{\n}\right)
\end{align}

Since the principal symbol of this operator (the highest order in derivatives) is not diagonal but it contains a non-minimal term given by the matrix $J_{AB}^{\a\b}$, the application of the Schwinger-DeWitt technique requires extra work.
\subsection{The Barvinsky-Vilkovisky technique}

There is a useful technique developed in \cite{Barvinsky} to compute the contribution to the quantum effective action of non-minimal operators. It is our aim to apply it to the case of $S_{UG}$ as given in \eqref{Sug}.

Let us concentrate on the highest derivative term of the operator 
\be
D_{AB}(\nabla)=\gamma_{AB}\Box+J_{AB}^{\a\b}\nabla_\a\nabla_\b
\ee
 Furthermore, assume that the full operator $F_{AB}$ can be included in a one-parameter family
\be \label{first_ass} F_{AB}(\nabla|\l)=D_{AB}(\nabla|\l)+M_{AB}.\hspace{1cm}0\leq \lambda\leq 1 \ee
so that $F_{AB}$ is minimal at $\lambda=0$. In our case, we can simply choose
\begin{align}\label{operator}
F_{AB}(\nabla|\lambda)=\gamma_{AB}\Box+\lambda J_{AB}^{\a\b}\nabla_\a\nabla_\b+M_{AB}
\end{align}
where $\lambda$ parametrizes the introduction of the non-minimal term in such a way that for $\l=0$ the operator is minimal and for $\l=1$ the operator whose determinant is desired is obtained.

Following Schwinger the effective action can be obtained by differentiating in $\lambda$ and integrating afterwards, arriving to\footnote{In order to simplify the notation, we are going to use a hat symbol for matrix operators carrying mixed capital indices. Thus, things like the following are assumed
\begin{align*}
&\hat{A}\equiv A=A_{A}^{B}\\
&\hat{A}\hat{B}\equiv AB=A_{A}^{B}B_{B}^{C}\\
&Tr(\hat{A})=tr \left(\gamma_{A}^{B}A_{B}^{A}\right)
\end{align*}
and so on and so forth. Here $tr$ denotes the usual matrix trace (i.e. sum of the elements of the diagonal).}
\be
 W(\l)=W(0)-\dfrac{1}{2}\int_0^\l d\l' Tr\left[\dfrac{d\hat{F}(\l')}{d\l'}\hat{G}(\l')\right]
\ee
where $\hat{G}(\lambda)$ is the Green function of the operator $F_{AB}$, defined by $\hat{F}(\l)\hat{G}(\l)=\mathbb{I}$, and we are dismissing ultralocal contributions and keeping only the pole part of $W(\lambda)$. The effective action for our original operator corresponds to $\lambda=1$. Here, $W(0)$ is the effective action of the corresponding minimal operator, obtained by setting $\lambda=0$.
\par
Many of the technical difficulties appear already in flat space. It is useful to consider the ordinary matrix in (euclidean) momentum space
\be
\hat{D}(k)=D^A_B(k)=\g^{AC}D_{CB}(k)
\ee 
with $k_\m$ a constant vector. Its inverse has the form,
\be \hat{D}^{-1}(k)=\dfrac{\hat{K}(k)}{(k^2)^m}\label{npower} \ee
with $m$ being an integer.

From this, it is clear that 
\be
\hat{D}(k)\hat{K}(k)=(k^2)^m\mathbb{I}.
\ee
 Were we to trade the vector $k_\m$ for the covariant derivative, and owing to the non-commutative character of the latter, a remainder appears
\be\hat{D}(\nabla)\hat{K}(\nabla)=\Box^m+\hat{K}_1(\nabla)\ee  
and going to the full operator, we get
\be \hat{F}(\nabla)\hat{K}(\nabla)=\Box^m+\hat{M}(\nabla) \ee
with $\hat{K}_1$ and $\hat{M}$ being now operators of as much order $2m-1$ in derivatives.

The last equation allows us to expand the Green function of $\hat{F}$ in powers of $\hat{M}$ as follows
\be\hat{G}=-\hat{K}\dfrac{\mathbb{I}}{\Box^m}\sum_{p=0}^4\left(-\hat{M}\dfrac{\mathbb{I}}{\Box^m}\right)^p+O \left(\mathfrak{m}^5\right)
\ee
The notation  $O \left(\mathfrak{m}^5\right)$ means that 
 we are keeping only terms up to background  dimension four.
 \par
 May be this is a good point to comment of power divergencies \cite{Fradkin,Alvarez} in the heat kernel formalism. It is possible to regularize the proper time integral in the Appendix \eqref{Wreg} by introducing both an ultraviolet $\Lambda^{-2}$ and infrared $\eta^{-2}$ cutoff, that is
 \be
 \int_0^\infty d\t\rightarrow \int_{\Lambda^{-2}}^{\eta^{-2}}
 \ee
 In that way we get
 \be
 W_\infty= {1\over 2} a_0 \Lambda^4+a_2 \Lambda^2+a_4\text{log}~\left({\Lambda\over \eta}\right)
 \ee
 It is to be stressed that those are gauge invariant {\em proper time cutoffs}, not to be confused with momentum cutoffs.
 \par
 The first term in this expansion yields  a universal (that is, independent of the form of the action) quartic renormalization of the cosmological constant. This term is non-dynamical in that it does not depend on any of the fields present in the theory. 
 \par
 The second term yields quadratic divergences, and the last term yields the physically most interesting ones, namely the logarithmic divergences. In pure gravity the quadratic divergences are necessarily proportional to
 \be
 \int R d\m(g)
 \ee
 which is the only dimension two invariant. The measure depends on whether full $\text{Diff}(M)$ invariance is implemented (as in GR, $d\m(g)\equiv \sqrt{|g|}d^n x$) or else only the subgroup $\text{TDiff}(M)$ (as in UG  $d\m(g)\equiv d^n x$). There has been some discussion going on in the literature on the physical relevance of those quadratic divergences {\em confer} \cite{Toms}\cite{Anber}\cite{Reuter}. 
 \par
 We shall concentrate on the logarithmic ones (which are the only ones seen in dimensional regularization) in the main body of the  paper and tackle the computation of the quadratic divergences in appendix \ref{app_D}.\par
In order to compute this Green function it is useful to commute the $\dfrac{\mathbb{I}}{\Box^m}$ to the right
\be\hat{G}=-\hat{K}\sum_{p=0}^4\left(-1\right)^p \hat{M}_p\dfrac{\mathbb{I}}{\Box^{m(p+1)}}+O\left(\mathfrak{m}^5\right)
\ee
with the operators $M_p$ given recursively by
\bea &&\hat{M}_0=\mathbb{I}\\
&&\hat{M}_{p+1}=\hat{M}\hat{M}_p+[\Box^m,\hat{M}_p]
\eea

Furthermore, it can be proven \cite{Barvinsky} that if the coefficient of the highest derivative term (of order $2d$) is covariantly conserved and there is no term of order $2d-1$, as it is the case for \eqref{operator}, then $M_4=M_3=0$ and $M_2=M^2+m[\Box,M]\Box^{m-1}$.

Turning now our attention to this explicit case and computing the inverse of the operator $\hat{D}(k)$ in the sense \eqref{npower} we get $m=3$ and the calculation of the effective action then reduces just to

\be 
W(\l)=W(0)-\dfrac{1}{2}\int_0^\l d\l' Tr\left[\hat{J}^{\a\b}\nabla_{\a}\nabla_{\b}\left\{\hat{K}\left(-\dfrac{\mathbb{I}}{\Box^3}+\hat{M}\dfrac{\mathbb{I}}{\Box^6}-3[\Box,\hat{M}]\dfrac{\mathbb{I}}{\Box^7}-\hat{M}^2\dfrac{\mathbb{I}}{\Box^9}\right)\right\}\right]\label{qeff}
\ee

The computation of $W(0)$ is just the one of a minimal second order operator cf.\cite{Barvinsky}
\begin{align}
W(0)=\dfrac{1}{16\pi^2}\frac{1}{n-4} \int d^{n}x\bigg\{ \frac{16}{15}R_{\m\n\a\b}R^{\m\n\a\b}+ \left(\frac{2}{8\a^{2}-1}-\frac{46}{15}\right)R_{\m\n}R^{\m\n}+\left(\frac{13}{24}+\frac{1}{2-16\a^2}\right)R^{2}\bigg\}
\end{align}

The rest of the pieces in \eqref{qeff} are obtained following the steps outlined above. The number of terms grows enormously after applying successive derivatives through Leibniz's rule. The computation has been performed with the help of the Mathematica software \emph{xAct} \cite{MartinGarcia}. A fair amount of computing time has been necessary in order to simplify the resulting expressions.

There is a last non-trivial issue that we have to take care of. After computing the terms in \eqref{qeff}, the output will be a collection of terms of the schematic form
\begin{align}
Tr\left({\cal O}_{\n_1\n_2...\n_j}\D_{\m_1}\D_{\m_2}...\D_{\m_p}\frac{\mathbb{I}}{\Box^n}\right)
\end{align}
with $p\leq 2n-4$.

These functional traces can be computed by introducing the formal representation of $\frac{\mathbb{I}}{\Box^n}$ through a Laplace transform and performing a dimensional regularization afterwards, keeping only the logarithmic divergent terms as explained in \cite{Barvinsky} and summarized in appendix \ref{app_C}.

After doing all this and computing the functional traces we are finally left with a simple result for the perturbation to $W(0)$
\begin{align}
\nonumber &-\dfrac{1}{2}\int_0^1 d\l' Tr\left[\hat{J}^{\a\b}\nabla_{\a}\nabla_{\b}\left\{\hat{K}\left(-\dfrac{\mathbb{I}}{\Box^3}+\hat{M}\dfrac{\mathbb{I}}{\Box^6}-3[\Box,\hat{M}]\dfrac{\mathbb{I}}{\Box^7}-\hat{M}^2\dfrac{\mathbb{I}}{\Box^9}\right)\right\}\right]=\\
&=\dfrac{1}{16\pi^2}\dfrac{1}{n-4}\int d^{n}x\;\bigg\{\left(\frac{1}{6 \alpha^2} + \frac{2}{1-8 \alpha^2}\right)R_{\m\n}R^{\m\n}+\frac{1}{24} \left(\frac{12}{8 \alpha^2 -1} - \frac{1}{\a^2}-5 \right)R^{2}\bigg\}
\end{align} 

And finally, putting all together we find that the contribution to the pole part of the effective action of $S_{UG}$ is
\begin{align}
W_{\infty}^{UG}=\dfrac{1}{16\pi^2}\frac{1}{n-4} \int d^{n}x\bigg\{\frac{16}{15}R_{\m\n\a\b}R^{\m\n\a\b}+\left(\frac{1}{6\a^2}-\frac{46}{15}\right)R_{\m\n}R^{\m\n}+\left(\frac{1}{3}-\frac{1}{24\a^2} \right)R^{2}\bigg\}
\end{align}
where we have neglected total derivatives in the integrand.

As has been already advertised, all the dependence on the gauge fixing, represented by the presence of the parameter $\alpha$ in the final result, disappears when we use the background equations of motion $R_{\m\n}=\frac{1}{4}R g_{\m\n}$. This is as it should be because all gauge fixing is BRST exact.

\subsection{The final result}
After computing the contribution of the non-minimal operator, we are finally ready to write the pole part of the effective action of Unimodular Gravity, which reads
\begin{align}
W_{\infty}=W_{\infty}^{UG}+W_{\infty}^{bc}+W_{\infty}^{\pi}+W_{\infty}^{\bar{c}\f}+W_{\infty}^{W}
\end{align}

Here $W_{\infty}^{UG}$ is the contribution we have computed in the last section while the rest of the contributions are given in appendix \ref{app_B}. Adding everything, we find that the final result is
\begin{align}
W_{\infty}=\dfrac{1}{16\pi^2}\frac{1}{n-4} \int d^{n}x\left(\frac{119}{90}R_{\m\n\a\b}R^{\m\n\a\b}+\left( \frac{1}{6 \alpha^2}-\frac{359}{90} \right) R_{\m\n}R^{\m\n}+\frac{1}{72} \left(22  - \frac{3}{\alpha^2}\right) R^{2}\right)
\end{align}

Now we would like to focus on the issue of on-shell renormalizability. It is known that although General Relativity is one-loop finite in the absence of a cosmological constant, this property is lost in its presence. The on-shell counterterm in this case was obtained in \cite{Christensen} and it amounts to a renormalization of the cosmological constant and is proportional to
\be
W^{GR}_\infty\equiv {1\over 16\pi^2 (n-4)}\int \sqrt{|g|} d^4 x~\left({53\over 45}~W_4-{1142\over 135}\Lambda^2\right)
\ee

Since the main attractive feature of Unimodular Gravity is precisely the different r\^ole that the cosmological constant plays with respect to GR, we would like to see what happens here with the renormalization group flow when we take the counterterm to be on-shell so that all external legs correspond to physical states. In that case, the equations of motion for the $|g|=1$ fixed background are the traceless Einstein equations
\begin{align}
R_{\m\n}-\frac{1}{4}R g_{\m\n}=0
\end{align}
which, altogether with Bianchi identities, imply the following for the operators appearing in the effective action
\begin{align}
&R_{\m\n\a\b}R^{\m\n\a\b}=E_{4}\\
&R_{\m\n}R^{\m\n}=\frac{1}{4}R^{2}\\
&R=\text{constant}
\end{align}
The first line is nothing more than the statement of the Gauss-Bonet theorem when we take into account the equations of motion. $E_4$ is thus Euler density, whose integral gives the Euler characteristic of the manifold.

By using these, we find that the on-shell effective action takes then the form
\begin{align}
W_{\infty}^{\text{on-shell}}=\dfrac{1}{16\pi^2}\frac{1}{n-4} \int d^{n}x\left(\frac{119}{90} E_{4}-\frac{83}{120} R^2\right)
\end{align}

The contribution of the cosmological constant to the effective action happens to be a non-dynamical quantity, since it does not couple to the metric because the $\sqrt{g}$ factor in the integration measure is absent. This implies that we can disregard this term since it will not contribute to any correlator involving physical fields. We conclude, therefore, that in this case there is no renormalization of the cosmological constant and its peculiar status in Unimodular Gravity is preserved through quantum corrections.

Indeed, this effect is not specific to one-loop computations.  We then conclude that the bare value of the cosmological constant is protected and quantum corrections do not modify it. 
\par
It could be thought that this effect is just a gauge artifact of our background choice $|\tilde{g}|=1$. However, it can be easily argued that this is not the case. As we have commented before in this work, if we now want to obtain the effective action for an arbitrary background from the one with unimodular background metric, it is enough to make a change of variables so that
\begin{align}
\tilde{g}_{\m\n}=g^{-\frac{1}{n}}g_{\m\n}
\end{align}

This transformation is available as long as there is no conformal anomaly in the theory. This is indeed the case, since there exists a regularization in which the anomaly vanishes \cite{Englert, Alvarezhv}
When doing this, we can see that the real reason of the cosmological constant not being renormalized is indeed the presence of Weyl invariance in our formalism, which protects the appearance of any mass scale in the effective action and, as a consequence, in the expectation value of the equations of motion. Therefore, our argument holds and the cosmological constant is protected and fixed to its bare value all along the renormalization group flow and at any loop order.

\section{Conclusions}
The cosmological constant problem appears in Unimodular Gravity in a different guise. The corresponding EM admit a first integral that plays the same role as the cosmological constant in General Relativity. The novelty is however that this first integral is not related to the zero momentum piece of the potential, but is rather determined by the boundary conditions, as is the rest of the dynamics. This is an important shift of the paradigm, in the sense that it explains why a huge value for the vacuum energy does not imply a correspondingly huge value for the cosmological constant.
\par
It has been argued in this paper that quantum corrections do not generate a cosmological constant in Unimodular Gravity. It would be more precise to say that the cosmological constant is generated, but it does not couple to the gravitational field. The analysis has been long and quite technical, but the result is simple enough. 
\par
It is worth pointing out that this result is not a consequence of the fact that we have chosen
\be
|\tilde{g}|=1
\ee
Unimodular gravity can be defined either by integrating under unimodular metrics only ${\cal D}\tilde{g}_{\a\b}$, or else under ${\cal D}\left( |g|^{-{1\over n}}~g_{\a\b}\right)$. In the latter case, there is a Weyl symmetry that forbids operators of zero dimension, that is
\be
S=M^n \int d^n x \left(-g\right)^\b
\ee
for any nonvanishing $\b$. This argument is independent of the background field technique, and holds in all regularizations that formally respect the symmetries of the bare action. From this point of view, what we have done is just to show that it is possible to implement one-loop quantum corrections in such a way that all symmetries of the classical action are respected.

 This is obviously independent of the gauge chosen in the background sector when using the background field technique in order to compute one loop results.

\par
We believe the present analysis is a step forward in the understanding of the cosmological constant.
\par
The physical content of Unimodular Gravity is quite similar to the one of General Relativity in spite of the technical complications caused by the absence of full $\emph{Diff}(M)$ symmetry and/or the presence of Weyl symmetry. It is worth exploring it in further detail to fully understand the subtle differences with General Relativity (which can always be worked out in the gauge $|g|=1 $ as Einstein himself did quite often).
\par
At any rate, the EM have to be worked out {\em before} gauge fixing. In General Relativity they are always of the form
\be
\sqrt{|g|}\left(\text{Something}\right)=\sqrt{|g|}~8\pi G~\left(\text{Something Else}\right)
\ee
and the two factors of $\sqrt{|g|}$ in both members cancel away, so that the unimodular gauge does not change the weight of vacuum energy in GR at the classical level. At the quantum level there is no symmetry to protect dimension zero terms to appear.
\par
Incidentally, no argument based on the EM is able to tell UG apart from GR. This is the case of the present formulation of string perturbation theory, where only on-shell amplitudes can be computed. This means that for the time being, UG is still a viable candidate for the low energy limit of string theory, at least up to tree level. We have not been able neither to show not to disprove that the full S-matrices are unitarily equivalent for both GR and UG to all orders in perturbation theory. We hope to come back to this fascinating topic soon.
\par
Let us finally stress that UG represents a Wilsonian solution to the conundrum pointed out in the Introduction; 
vacuum energy is effectively weightless at a minimal cost. Neither the equivalence principle nor minimal coupling are in need of modification.

\section{Acknowledgments}
We are grateful for illuminating discussions with Andrei Barvinsky and Christian Steinwachs. This work has been partially supported by the European Union FP7 ITN INVISIBLES (Marie Curie Actions, PITN- GA-2011- 289442)and (HPRN-CT-200-00148); COST action MP1405 (Quantum Structure of Spacetime), COST action MP1210 (The String Theory Universe) as well as by FPA2012-31880 (MICINN, Spain)), FPA2011-24568 (MICINN, Spain), and S2009ESP-1473 (CA Madrid). The authors acknowledge the support of the Spanish MINECO {\em Centro de Excelencia Severo Ochoa} Programme under grant SEV-2012-0249. 
The xAct package \cite{MartinGarcia} has been extensively used in the computations of the present paper.

\newpage
\appendix
\section{Renormalized one-loop effective action from the Heat Kernel}\label{heat_kernel}

The most powerful technique for computing the pole part of the renormalized effective action in curved space is the Heat Kernel or Schwinger-DeWitt method \cite{DeWitt}, which is based upon the construction of the following functional trace
\begin{align}
K(s,f,D)=Tr\left(f\;e^{-sD}\right)
\end{align}

The quantum effective action of a quantum field theory can be written in the one-loop approximation and in the absence of external sources as
\begin{align*}
W=-\frac{1}{2}\log\left(\det~D\right)
\end{align*}
where $D$ is the second order operator governing one-loop fluctuations. 

This determinant is divergent and must be regularized. Consider a suitable set of eigenfunctions $\psi_{i}$ of $D$ such that each one corresponds to an eigenvalue $\lambda_{i}$. Thus, we can define the zeta function $\zeta(s,D)$ of the operator $D$ as
\begin{align}
\zeta(s,D)=\Gamma(s)^{-1} \int_{0}^{\infty}dt\; t^{s-1} \left<\psi_{i}\right|e^{-tD}\left|\psi_{i}\right>=\int_{0}^{\infty}dt\; t^{s-1} K(t,D)
\end{align}
where $K(t,D)=Tr\left(e^{-tD}\right)$ is known to be \emph{traced Heat Kernel} of the operator $D$. By using this we can formally represent $\log\left(det(D)\right)$ as
\begin{align}
\log\left(\text{det}~D\right)=-\int^{\infty}_{0}\frac{dt}{t} \left<\psi_{i}\right|e^{-tD}\left|\psi_{i}\right>=-\int^{\infty}_{0}\frac{dt}{t}K(t,D)
\end{align}
so that the one-loop effective action is
\begin{align}\label{effective_action_hk}
W=-\frac{1}{2}\int^{\infty}_{0}\frac{dt}{t}K(t,D)
\end{align}

Finally, we shift the power of $t$ and introduce the zeta function as a regularization scheme, having
\begin{align}\label{Wreg}
W_{reg}=-\frac{1}{2}\mu^{2s}\int_{0}^{\infty}\frac{dt}{t^{1-s}}K(t,D)=-\frac{1}{2}\mu^{2s}\Gamma(s)\zeta(s,D)
\end{align}
where $\mu$ is a constant of the dimension of mass that will play the role of the usual scale in dimensional regularization.

Now, in order to remove the regularization scheme and define the renormalized theory, we have to take the limit $s\rightarrow 0$. The regularized effective action has a pole around this value
\begin{align}
W_{reg}=-\frac{1}{2}\left(\frac{1}{s}-\gamma_{E}+\log\left(\mu^{2}\right)\right)\zeta(0,D)-\frac{1}{2}\zeta^{'}(0,D)
\end{align}
with $\gamma_{E}$ being the Euler-Mascheroni constant.

This pole term has to be removed by renormalization and, by using a minimal subtraction scheme, the remaining part of $W_{reg}$ will be the renormalized action of the theory
\begin{align}
W_{ren}=-\frac{1}{2}\zeta^{'}(0,D)-\frac{1}{2}\log\left(\mu^{2}\right)\zeta(0,D)
\end{align}
where we have rescaled the constant $\mu$ to absorb $\gamma_{E}$. Thus, the one-loop logarithmic divergences of the effective action are encoded in the $s\rightarrow 0$ limit of $\zeta(s,D)$.

The interesting point is that this limit is encoded in the asymptotic expansion of the Heat Kernel when $s\rightarrow0$. In this limit, the Heat Kernel behaves as
\begin{align}
K(s,D)=\frac{1}{(4 \pi s)^{n/2}}\sum_{i=0}s^{i/2}a_{i}(D)
\end{align}
where the coefficients $a_{i}(t,D)$ are computable in terms of local invariants of the manifold. In particular, it can be shown that in the physical dimension, $d$ (which for us will be d=4)
\begin{align}
\zeta(0,D)=a_{d}(D)
\end{align}
so that for a field theory in $d$ dimensions, the pole part of the one-loop effective action is given by
\begin{align}
W_{\infty}=\dfrac{1}{(4\pi)^{n/2}}\frac{1}{n-d} a_{d}(D)
\end{align}
where dimensional regularization around $n=d$ has been used in the last step.

To summarize, the computation of the pole part of the quantum effective action reduces to the computation of the $a_{n}(D)$ heat kernel coefficient of a given operator.

\subsection{Heat Kernel coefficients of a Laplace-type operator}

Most of the operators involved in the computation of the effective action of Unimodular Gravity are second order Laplace-type operators of the form
\begin{align}\label{Laplace operator}
D=-\g_{AB}\Box+N^{\m}_{AB}\D_{\m}+M_{AB}
\end{align}
where the capital indices refer to some possible gauge bundle. It can be always taken to a simple form after redefining the covariant derivative as ${\cal D}=\D+\omega$, so that
\begin{align}
D=-\gamma_{AB}{\cal D}^{2}-E_{AB}
\end{align}
where 
\begin{align}
&\omega_{\m\;B}^{A}=\frac{1}{2}\g^{AC}N_{\m CB}\\
&E^{A}_{B}=\g^{AC}(-M_{CB}-\omega_{\m CF}\omega^{\m\;F}_{B}-\D_{\m}\omega^{\m}_{CB})
\end{align}

Once in this form, the Heat Kernel expansion for such operator has been computed in \cite{DeWitt}. The relevant coefficient for $n=4$ reads
\begin{equation}
 \label{a4}a_{4}=\frac{1}{360}\int d^{n}x\sqrt{|g|}\;Tr(60\Box E +60 RE + 180 E^2 +12\Box R + 5R^{2}-2R_{\m\n}R^{\m\n}+2R_{\m\n\r\s}R^{\m\n\r\s}+30\hat{\cal{R}}_{\m\n}\hat{\cal{R}}^{\m\n})
\end{equation}
which correspond to the quantum effective action for four dimensions. We have included total derivatives since they contribute if one is interested in the phenomenon of conformal anomalies. Here $\hat{\cal{R}}_{\m\n}$ refers to the field strength defined by
\begin{align}\label{field_strength}
[{\cal D}_{\m},{\cal D}_{\n}]V_{A}=\hat{\cal{R}}_{\m\n A}^{\;\;\;\;\;\;B}V_{B}
\end{align}
Concerning the graviton fluctuations this is given by the Ricci identity.
\bea
[\nabla_\m ,\nabla_\n]h^{\r\s}&&=\hat{\cal R}^{\r\s}_{\m\n\a\b}h^{\a\b}\\
\hat{\cal R}^{\r\s}_{\m\n\a\b}&&=\dfrac{1}{2}\left(R^\r_{\a\m \n}\d^\s_\b+R^\s_{\a\m\n}\d^\r_\b+R^\r_{\b\m\n}\d^\s_\a+R^\s_{\b\m\n}\d^\r_\a\right)
\eea
As a special case to be considered later, let us take the simplest possible operator
\begin{align}
D=-\Box
\end{align}
acting onto a scalar field. Here both $N^{\m}$ and $M$ are zero and the field strength vanishes. This means that its Heat Kernel $a_{4}$ coefficient is just
\begin{align}\label{a4box}
a_{4}(\Box)=\frac{1}{360}\int d^{n}x\sqrt{|g|}\;\left(12\Box R + 5R^{2}-2R_{\m\n}R^{\m\n}+2R_{\m\n\r\s}R^{\m\n\r\s}\right)
\end{align}

\subsection{Heat Kernel coefficients of a quartic operator}

One of the operators appearing in our computations is an operator whose leading part contains four covariant derivatives. The Heat Kernel of these operators have been also studied by many people and fairly general formulas have been given. However, here we are only interested in the contribution to the effective action in four dimensions. This has been computed in \cite{Barvinsky, Gusynin} for an operator of the form we are interested in
\begin{align}\label{quartic_operator}
D=\gamma_{AB}\Box^{2}+\Omega_{AB}^{\m\n\a}\D_{\m}\D_{\n}\D_{\a}+J_{AB}^{\m\n}\D_{\m}\D_{\n}+H_{AB}^{\m}\D_{\m}+P_{AB}
\end{align}

The corresponding expression of this kind of operators in four dimensions is quite involved. However, when $\Omega^{\m\n\a}_{AB}=0$ as it is in the case of our work, the resultant expression simplifies a lot and reads, with our conventions
\begin{align}\label{a4_quartic}
\nonumber W_{\infty}=\dfrac{1}{16\pi^2}\frac{1}{n-4} \int& d^{n}x \sqrt{|g|}\;Tr\left(\frac{1}{90}R_{\m\n\a\b}R^{\m\n\a\b}-\frac{1}{90}R_{\m\n}R^{\m\n}+\frac{1}{36}R^{2}\mathbb{I}- \hat{P}+\frac{1}{6}\hat{\cal{R}}_{\m\n}\hat{\cal{R}}^{\m\n}-\right.\\
&\left.-\frac{1}{6}J^{(\m\n)}R_{\m\n}+\frac{1}{12} J_{\m}^{\m} R+\frac{1}{48}(J^{\m}_{\m})^{2}+\frac{1}{24}J_{(\m\n)}J^{(\m\n)}-\dfrac{1}{2}J^{[\m\n]}\hat{\cal{R}}_{\m\n}\right)
\end{align}

where, as usual
\begin{align}J^{(\m\n)}&=\dfrac{1}{2}\left(J^{\m\n}+J^{\n\m}\right)\\
J^{[\m\n]}&=\dfrac{1}{2}\left(J^{\m\n}-J^{\n\m}\right)
\end{align}

\section{Heat Kernel contributions of the different operators involved}\label{app_B}
Here we compute the different heat kernel coefficients corresponding to each of the minimal differential operator appearing in the path integral formulation of Unimodular Gravity.

\subsection{The contribution of $S_{bc}$}
The action term for the fields $b^{\m(1,-1)}$ and $c^{\m(1,1)}$ was defined in equation \eqref{Sbc} and reads
\begin{align}
\int d^{n}x\;b^{\m}&\left(\Box^{2}c^{(1,1)}_{\m}-2R_{\m\r}\D^{\r}\D_{\n}c^{\n(1,1)}-\Box R_{\m\r}c^{\r(1,1)}-2\D_{\s}R_{\m\r}\D^{\s}c^{\r(1,1)}-R_{\m\r}R^{\r\n}c_{\n}^{(1,1)}\right)
\end{align}

This is a quartic operator of the form \eqref{quartic_operator} if we identify
\begin{align}
&J^{\m\n}_{\a\b}=-2R^{\m}_{\a}\delta^{\n}_{\b}\\
&H^{\m}_{\a\b}=-2\D^{\m}R_{\a\b}\\
&P_{\a\b}=-\Box R_{\a\b}-R_{\a\r}R^{\r}_{\b}
\end{align}
here the bundle indices are just spacetime greek indices that we indicate with $\a$ and $\b$.

And the field strength
\begin{align}
[\D_{\m},\D_{\n}]c^{\a}=\hat{\cal{R}}_{\m\n}^{\;\;\;\a\b}c_{\b}=R_{\m\n}^{\;\;\;\a\b}c_{\b}
\end{align}

Plugging this into \eqref{a4_quartic} we find that the contribution of $S_{bc}$ to the quantum effective action is
\begin{align}
W_{\infty}^{bc}=\dfrac{1}{16\pi^2}\frac{1}{n-4} \int d^{n}x\left(\frac{11}{45}R_{\m\n\a\b}R^{\m\n\a\b}-\frac{41}{45}R_{\m\n}R^{\m\n}-\frac{1}{18}R^{2}\right)
\end{align}
where we have set $|g|=1$ and we have multiplied by minus two in order to take into account of the fact that there are two fermionic fields.

\subsection{The contribution of $S_{\pi}$}
The action term for the dynamics of the fermionic $\pi$ fields was defined in \eqref{Spi} and reads
\begin{align}
S_{\pi}=\int d^{n}x\;\pi^{(1,-1)}\Box^{-1}\pi'^{\;(1,1)}
\end{align}

Even if this is a pseudodifferential operator, its contribution to the pole part of the quantum effective action can be easily computed thanks to the fact that $\Box \times \Box^{-1}=1$. This means that 
\begin{align}
Det(\Box)=Det(\Box^{-1})^{-1}\longrightarrow \log\left[Det(\Box)\right]=-\log\left[Det(\Box^{-1})\right]
\end{align}
if there is no multiplicative anomaly. This sums up into the fact that the corresponding Heat Kernel expansion of $\Box^{-1}$ will be minus the expansion of $\Box$. Therefore, by using the result of \eqref{a4box}
\begin{align}
a_{4}(\Box)=-\frac{1}{360}\int d^{n}x\;\left(12\Box R + 5R^{2}-2R_{\m\n}R^{\m\n}+2R_{\m\n\r\s}R^{\m\n\r\s}\right)
\end{align}
where we have already set $g=1$.

However, here we are integrating over two fermionic fields, which introduces another factor of minus two. Thus, we have that
\begin{align}
a_{4}^{\pi}=\frac{1}{180}\int d^{n}x\;\left(12\Box R + 5R^{2}-2R_{\m\n}R^{\m\n}+2R_{\m\n\r\s}R^{\m\n\r\s}\right)
\end{align}
and its contribution to the effective action is given by
\begin{align}
W_{\infty}^{\pi}=\dfrac{1}{16\pi^2}\frac{1}{n-4} \frac{1}{180}\int d^{n}x\;\left(12\Box R + 5R^{2}-2R_{\m\n}R^{\m\n}+2R_{\m\n\r\s}R^{\m\n\r\s}\right)
\end{align}

\subsection{The contribution of $S_{\bar{c}\f}$}
The action term for $\bar{c}$ and $\f$ was given in equation \eqref{Scphi}, reading
\begin{align}
\int d^{n}x\; \bar{c}^{(0,-2)} \Box \f^{(0,2)}
\end{align}

This is the simplest possible operator and its $a_{4}$ coefficient was given in \eqref{a4box}. It reads
\begin{align}
a_{4}^{\bar{c}\f}=\frac{1}{180}\int d^{n}x\;\left(12\Box R + 5R^{2}-2R_{\m\n}R^{\m\n}+2R_{\m\n\r\s}R^{\m\n\r\s}\right)
\end{align}
where a factor of two has been introduced to take into account that we have two fields. Again, remind that we have set $g=1$.

Its contribution to the effective action is given by
\begin{align}
W_{\infty}^{\bar{c}\f}=\dfrac{1}{16\pi^2}\frac{1}{n-4} \frac{1}{180}\int d^{n}x\;\left(12\Box R + 5R^{2}-2R_{\m\n}R^{\m\n}+2R_{\m\n\r\s}R^{\m\n\r\s}\right)
\end{align}

\subsection{The contribution of $S_{W}$}
The action term for the Weyl ghost field was given in \eqref{SW} and reads
\begin{align}
2n\alpha\int d^{n}x\; b \Box c
\end{align}

The global multiplicative constant will not contribute to the pole part of the quantum effective action, since it gives just an ultralocal contribution, so we can dismiss it, having just
\begin{align}
\int d^{n}x\; b \Box c
\end{align}

Again, we are left the simplest possible operator and its $a_{4}$ coefficient was given in \eqref{a4box}. It reads
\begin{align}
a_{4}^{W}=-\frac{1}{180}\int d^{n}x\;\left(12\Box R + 5R^{2}-2R_{\m\n}R^{\m\n}+2R_{\m\n\r\s}R^{\m\n\r\s}\right)
\end{align}
where a factor of minus two has been introduced to take into account that we have two fermionic fields and we have set again $g=1$.

Its contribution to the effective action is given by
\begin{align}
W_{\infty}^{W}=-\dfrac{1}{16\pi^2}\frac{1}{n-4} \frac{1}{180}\int d^{n}x\;\left(12\Box R + 5R^{2}-2R_{\m\n}R^{\m\n}+2R_{\m\n\r\s}R^{\m\n\r\s}\right)
\end{align}

\section{Functional traces}\label{app_C}
The functional traces 
\begin{align}
Tr\left({\cal O}_{\n_1\n_2...\n_j}\D_{\m_1}\D_{\m_2}...\D_{\m_p}\frac{\mathbb{I}}{\Box^n}\right) \label{trazas}
\end{align}
that appear in the calculation of the quantum effective action will lead to new contributions to the divergences and can be computed using the heat kernel representation of the operator.

Starting with an operator $\hat{F}(\nabla)$, it can be written as\footnote{Let us note here that while \cite{Barvinsky} are performing their computations in lorentzian signature, we are doing them in the euclidean setting. The differences account for some global signs and some factors of $i$ in the definition of the proper time and the effective action.}	
\be (\hat{F}(\nabla))^{-n}=\dfrac{1}{(n-1)!}\left[\left(\dfrac{d}{dm^2}\right)^{n-1}G(m^2)\right]_{m^2=0}\ee

Now, the heat kernel representation of the Green function is
\be G(m^2)=\int^\infty_0 \exp{(- s m^2)}\exp{(-s \hat{F}(\nabla))}\ee
where
\be \exp(-s\hat{F}(\nabla))\delta(x,x')=\dfrac{1}{(4\pi )^{n/2}}\dfrac{{\cal{D}} ^{1/2} (x,x')}{s^{n/2}}\exp{\left(-\dfrac{\sigma(x,x')}{2s}\right)}\hat{\O}(s|x,x')\ee
and with
\be\hat{\O}(s|x,x')=\sum_{n=0}^\infty s^n \hat{a}_n(x,x')\ee

Here $\sigma$ is the world function, defined by the equation $\sigma=\dfrac{1}{2}\sigma_\m\sigma^\m$ and ${\cal{D}}(x,x')$ is the so-called Van-Vleck determinant
\bea {\cal{D}}(x,x')&=&\left|det\left(-\dfrac{\partial\sigma}{\partial x^\m\partial x'^\n}\right)\right|\\
{\cal{D}}(x,x')&=& g^{1/2}(x)g^{1/2}(x')\Delta(x,x')
\eea

For the particular case of $\hat{F}(\nabla)=\hat{\Box}$ we can find the representation of the inverse Laplace operator
\be \dfrac{\mathbb{I}}{\Box^n}=\dfrac{1}{(n-1)!}\int_0^\infty ds\; s^{n-1}\exp{(-s\hat{\Box})}\ee

Each of the traces we find in our computation can now be computed by acting with derivatives on this representation and using the tables of coincidence limits given in \cite{Barvinsky}. Finally it is needed to integrate over $s$, where we find that only three types of (logarithmic) divergent integrals arise for dimension $n\rightarrow 4$
\be \int_0^\infty \dfrac{ds}{s^{n/2+k}}, \;\;\text{with}\;\;\; k=-1,0,1
\ee
and whose pole part can be obtained by integrating by parts, which gives the Laurent series of the result.

All but one of the functional traces we need in our computation can be found in \cite{Barvinsky}. Here we give the value of all of them and remark that we have rederived all of them explicitly, thus checking their results.

The divergent functional traces corresponding to p=2n \eqref{trazas} that appear are
\begin{eqnarray}
	\nabla_\m\nabla_\n\nabla_\a\nabla_\b\dfrac{\mathbb{I}}{\Box^2}&=&\dfrac{\sqrt{g}}{8(n-4)\pi^2}\left\{ \left[\dfrac{1}{36}\left(R_{\m\n}R_{\a\b}+R_{\m\a}R_{\n\b}+R_{\m\b}R_{\n\a}\right)+\dfrac{1}{180}\left(R_\m^\l\left(11R_{\n\a\b\l}-R_{\b\a\n\l}\right)+\right.\right.\right.\nonumber\\&& \left.\left.\left.+R_\n^\l\left(11R_{\m\a\b\l}-R_{\b\a\m\l}\right)+R_\a^\l\left(11R_{\m\n\b\l}-R_{\b\n\m\l}\right)+R_\b^\l\left(11R_{\m\n\a\l}-R_{\a\n\m\l}\right)\right)+\right.\right.\nonumber\\&& \left.\left.+\dfrac{1}{90}\left(R_{\m\;\;\n}^{\;\;\l\;\;\s}\left(R_{\l\a\s\b}+R_{\l\b\s\a}\right)+R_{\m\;\;\a}^{\;\;\l\;\;\s}\left(R_{\l\n\s\b}+R_{\l\b\s\n}\right)+R_{\m\;\;\b}^{\;\;\l\;\;\s}\left(R_{\l\n\s\a}+R_{\l\a\s\n}\right)\right)+\right.\right.\nonumber\\&&\left.\left. +\dfrac{1}{20}\left(\nabla_\m\nabla_\n R_{\a\b}+\nabla_\m\nabla_\a R_{\n\b}+\nabla_\m \nabla_\b R_{\n\a}+\nabla_n\nabla_a R_{\m\b}+\nabla_\n\nabla_\b R_{\m\a}
+\nabla_a\nabla_\b R_{\m\n}\right)\right]\mathbb{I}+\right.\nonumber\\&&\left.+\dfrac{1}{12}\left[R_{\m\n}\hat{\cal{R}}_{\a\b}+R_{\m\a}\hat{\cal{R}}_{\n\b}+R_{\m\b}\hat{\cal{R}}_{\n\a}+R_{\n\a}\hat{\cal{R}}_{\m\b}+R_{\n\b}\hat{\cal{R}}_{\m\a}+R_{\a\b}\hat{\cal{R}}_{\m\n}\right]+\right.\nonumber\\&&\left.+\dfrac{1}{2}\left[\nabla_\m\nabla_\n\hat{\cal{R}}_{\a\b}+\nabla_\m\nabla_\a\hat{\cal{R}}_{\n,b}+\nabla_\n\nabla_\a\hat{\cal{R}}_{\m\b}\right]+\dfrac{1}{8}\left[\hat{\cal{R}}_{\m\n}\hat{\cal{R}}_{\a\b}+\hat{\cal{R}}_{\a\b}\hat{\cal{R}}_{\m\n}+\hat{\cal{R}}_{\m\a}\hat{\cal{R}}_{\n\b}+\right.\right.\nonumber\\&&\left.\left.+\hat{\cal{R}}_{\n\b}\hat{\cal{R}}_{\m\a}+\hat{\cal{R}}_{\m\b}\hat{\cal{R}}_{\n\a}+\hat{\cal{R}}_{\n\a}\hat{\cal{R}}_{\m\b}\right]-\dfrac{1}{12}\left[\hat{\cal{R}}_{\m\l}\left(R_{\a\n\b}^\l+R_{\b\n\a}^\l\right)+\hat{\cal{R}}_{\n\l}\left(R_{\a\m\b}^\l+R_{\b\m\a}^\l\right)+\right.\right.\nonumber\\&&\left.\left.+\hat{\cal{R}}_{\a\l}\left(R_{\n\m\b}^\l+R_{\b\m\n}^\l\right)+\hat{\cal{R}}_{\b\l}\left(R_{\m\n\a}^\l+R_{\a\n\m}^\l\right)\right]-\dfrac{1}{2}\left[-\dfrac{1}{9}\left(R_{\a\m\b\n}+R_{\b\m\a\n}\right)R\mathbb{I}+\right.\right.\nonumber\\
&&\left.\left.+g_{\m\n}\left[\left(\dfrac{1}{36}R_{\a\b}R+\dfrac{1}{90}R^{\l\s}R_{\l\a\s\b}+\dfrac{1}{90}R_{\r\s\l\a}R^{\r\s\l}_{\;\;\;\;\;\b}-\dfrac{1}{45}R_{\a\l}R^\l_\b+\dfrac{1}{60}\Box R_{\a\b}+\dfrac{1}{20}\nabla_\a\nabla_\b R\right)\mathbb{I}+\right.\right.\right.\nonumber\\&&\left.\left.\left.+ \dfrac{1}{12}\left(\hat{\cal{R}}_{\a\l}\hat{\cal{R}}^\l_{\;\b}+\hat{\cal{R}}_{\b\l}\hat{\cal{R}}^\l_{\;\a}\right)-\dfrac{1}{12}\left(\nabla_\a\nabla^\l\hat{\cal{R}}_{\l\b}+\nabla_\b\nabla^\l\hat{\cal{R}}_{\l\a}\right)+\dfrac{1}{12}R \hat{\cal{R}}_{\a\b}\right]+\right.\right.\nonumber\\
&&\left.\left.+g_{\m\a}\left[\left(\dfrac{1}{36}R_{\n\b}R+\dfrac{1}{90}R^{\l\s}R_{\l\n\s\b}+\dfrac{1}{90}R_{\r\s\l\n}R^{\r\s\l}_{\;\;\;\;\;\b}-\dfrac{1}{45}R_{\n\l}R^\l_\b+\dfrac{1}{60}\Box R_{\n\b}+\dfrac{1}{20}\nabla_\n\nabla_\b R\right)\mathbb{I}+\right.\right.\right.\nonumber\\&&\left.\left.\left.+ \dfrac{1}{12}\left(\hat{\cal{R}}_{\n\l}\hat{\cal{R}}^\l_{\;\b}+\hat{\cal{R}}_{\b\l}\hat{\cal{R}}^\l_{\;\n}\right)-\dfrac{1}{12}\left(\nabla_\n\nabla^\l\hat{\cal{R}}_{\l\b}+\nabla_\b\nabla^\l\hat{\cal{R}}_{\l\n}\right)+\dfrac{1}{12}R \hat{\cal{R}}_{\n\b}\right]+\right.\right.\nonumber\\
&&\left.\left.+g_{\m\b}\left[\left(\dfrac{1}{36}R_{\n\a}R+\dfrac{1}{90}R^{\l\s}R_{\l\n\s\a}+\dfrac{1}{90}R_{\r\s\l\n}R^{\r\s\l}_{\;\;\;\;\;\a}-\dfrac{1}{45}R_{\n\l}R^\l_\a+\dfrac{1}{60}\Box R_{\n\a}+\dfrac{1}{20}\nabla_\n\nabla_\a R\right)\mathbb{I}+\right.\right.\right.\nonumber\\&&\left.\left.\left.+ \dfrac{1}{12}\left(\hat{\cal{R}}_{\n\l}\hat{\cal{R}}^\l_{\;\a}+\hat{\cal{R}}_{\a\l}\hat{\cal{R}}^\l_{\;\n}\right)-\dfrac{1}{12}\left(\nabla_\n\nabla^\l\hat{\cal{R}}_{\l\a}+\nabla_\a\nabla^\l\hat{\cal{R}}_{\l\n}\right)+\dfrac{1}{12}R \hat{\cal{R}}_{\n\a}\right]+\right.\right.\nonumber\\
&&\left.\left.+g_{\n\a}\left[\left(\dfrac{1}{36}R_{\m\b}R+\dfrac{1}{90}R^{\l\s}R_{\l\m\s\b}+\dfrac{1}{90}R_{\r\s\l\m}R^{\r\s\l}_{\;\;\;\;\;\b}-\dfrac{1}{45}R_{\m\l}R^\l_\b+\dfrac{1}{60}\Box R_{\m\b}+\dfrac{1}{20}\nabla_\m\nabla_\b R\right)\mathbb{I}+\right.\right.\right.\nonumber\\&&\left.\left.\left.+ \dfrac{1}{12}\left(\hat{\cal{R}}_{\m\l}\hat{\cal{R}}^\l_{\;\b}+\hat{\cal{R}}_{\b\l}\hat{\cal{R}}^\l_{\;\m}\right)-\dfrac{1}{12}\left(\nabla_\m\nabla^\l\hat{\cal{R}}_{\l\b}+\nabla_\b\nabla^\l\hat{\cal{R}}_{\l\m}\right)+\dfrac{1}{12}R \hat{\cal{R}}_{\m\b}\right]+\right.\right.\nonumber\\
&&\left.\left.+g_{\n\b}\left[\left(\dfrac{1}{36}R_{\m\a}R+\dfrac{1}{90}R^{\l\s}R_{\l\m\s\a}+\dfrac{1}{90}R_{\r\s\l\m}R^{\r\s\l}_{\;\;\;\;\;\a}-\dfrac{1}{45}R_{\m\l}R^\l_\a+\dfrac{1}{60}\Box R_{\m\a}+\dfrac{1}{20}\nabla_\m\nabla_\a R\right)\mathbb{I}+\right.\right.\right.\nonumber\\&&\left.\left.\left.+ \dfrac{1}{12}\left(\hat{\cal{R}}_{\m\l}\hat{\cal{R}}^\l_{\;\a}+\hat{\cal{R}}_{\a\l}\hat{\cal{R}}^\l_{\;\m}\right)-\dfrac{1}{12}\left(\nabla_\m\nabla^\l\hat{\cal{R}}_{\l\a}+\nabla_\a\nabla^\l\hat{\cal{R}}_{\l\m}\right)+\dfrac{1}{12}R \hat{\cal{R}}_{\m\a}\right]+\right.\right.\nonumber\\
&&\left.\left.+g_{\a\b}\left[\left(\dfrac{1}{36}R_{\m\n}R+\dfrac{1}{90}R^{\l\s}R_{\l\m\s\n}+\dfrac{1}{90}R_{\r\s\l\m}R^{\r\s\l}_{\;\;\;\;\;\n}-\dfrac{1}{45}R_{\m\l}R^\l_\n+\dfrac{1}{60}\Box R_{\m\n}+\dfrac{1}{20}\nabla_\m\nabla_\n R\right)\mathbb{I}+\right.\right.\right.\nonumber\\&&\left.\left.\left.+ \dfrac{1}{12}\left(\hat{\cal{R}}_{\m\l}\hat{\cal{R}}^\l_{\;\n}+\hat{\cal{R}}_{\n\l}\hat{\cal{R}}^\l_{\;\m}\right)-\dfrac{1}{12}\left(\nabla_\m\nabla^\l\hat{\cal{R}}_{\l\n}+\nabla_\n\nabla^\l\hat{\cal{R}}_{\l\m}\right)+\dfrac{1}{12}R \hat{\cal{R}}_{\m\n}\right]\right]+\right.\nonumber\\&&\left. +\dfrac{1}{4}\left(g_{\m\n}g_{\a\b}+g_{\m\a}g_{\n\b}+g_{\m\b}g_{\n\a}\right)\left[\left[\dfrac{1}{180}\left(R_{\l\s\r\g}R^{\l\s\r\g}-R_{\l\d}R^{\l\s}\right)+\dfrac{1}{30}\Box R-\dfrac{1}{72}R^2\right]\mathbb{I}+\right.\right.\nonumber\\&&\left.\left.+\dfrac{1}{12}\hat{\cal{R}}_{\l\s}\hat{\cal{R}}^{\l\s}\right]\right\}
\end{eqnarray}

\begin{eqnarray}
	\nabla_\m\nabla_\n\dfrac{\mathbb{I}}{\Box}&=& \dfrac{\sqrt{g}}{8(n-4)\pi^2}\dfrac{1}{2}\left\{\left[-g_{\m\n}\left(\dfrac{1}{180}R_{\a\b\l\s}R^{\a\b\l\s}-\dfrac{1}{180}R_{\a\b}R^{\a\b}+\dfrac{1}{72}R^2+\dfrac{1}{30}\Box R\right)\mathbb{I}+\right.\right.\nonumber\\&&\left.+\dfrac{1}{45}R^{\a\b}R_{\a\m\b\n}+\dfrac{1}{45}R_{\a\b\l\m}R^{\a\b\l}_\n-\dfrac{2}{45}R_{\m\a}R^\a_\n+\dfrac{1}{18}R R_{\m\n}+\dfrac{1}{30}\Box R_{\m\n}\right.+\nonumber\\&&\left.+\dfrac{1}{10}\nabla_\m\nabla_\n R\right]-\dfrac{1}{12}g_{\m\n}\hat{\cal{R}}_{\a\b}\hat{\cal{R}}^{\a\b}+\dfrac{1}{6}R\hat{\cal{R}}_{\m\n}+\dfrac{1}{6}\hat{\cal{R}}_{\m\a}\hat{\cal{R}}_\n^\a+\dfrac{1}{6}\hat{\cal{R}}_{\n\a}\hat{\cal{R}}_\m^\a\nonumber -\\&&-\dfrac{1}{6}\nabla_\m\nabla^\a\hat{\cal{R}}_{\a\n}-\dfrac{1}{6}\nabla_\n\nabla^\a\hat{\cal{R}}_{\a\m}\bigg\}
\end{eqnarray}

For $p=2n-1$ just one is involved

\begin{flalign}
	&\nabla_\m\dfrac{\mathbb{I}}{\Box}= \dfrac{\sqrt{g}}{8(n-4)\pi^2}\left(\dfrac{1}{12}\nabla_\m R\mathbb{I}-\dfrac{1}{6}\nabla^\n\hat{\cal{R}}_{\n\m}\right)&
\end{flalign}

The ones with $p=2n-2$

\begin{flalign}& \dfrac{\mathbb{I}}{\Box}=\dfrac{\sqrt{g}}{8(n-4)\pi^2}\dfrac{1}{6}R\mathbb{I}&
\end{flalign}

\begin{flalign}
	&\nabla_\m\nabla_\n\dfrac{\mathbb{I}}{\Box^2}= -\dfrac{\sqrt{g}}{8(n-4)\pi^2}\left[\dfrac{1}{6}\left(R_{\m\n}-\dfrac{1}{2}g_{\m\n}R\right)\mathbb{I}+\dfrac{1}{2}\hat{\cal{R}}_{\m\n}\right]&
\end{flalign}

\begin{eqnarray}
	\nabla_\a\nabla_\b\nabla_\m\nabla_\n\dfrac{\mathbb{I}}{\Box^3}&=& -\dfrac{\sqrt{g}}{8(n-4)\pi^2}\dfrac{1}{4}\left\{-\dfrac{2}{6}R_{\m\b\n\a}-\dfrac{2}{6}R_{\n\b\m\a}\mathbb{I}+g_{\m\n}\left(\dfrac{1}{6}R_{\a\b}\mathbb{I}+\dfrac{1}{2}\hat{\cal{R}}_{\a\b}\right)+\right.\nonumber\\ &&+\left.g_{\b\n}\left(\dfrac{1}{6}R_{\a\m}\mathbb{I}+\dfrac{1}{2}\hat{\cal{R}}_{\a\m}\right)+ g_{\a\n}\left(\dfrac{1}{6}R_{\m\b}\mathbb{I}+\dfrac{1}{2}\hat{\cal{R}}_{\m\b}\right)+g_{\b\m}\left(\dfrac{1}{6}R_{\a\n}\mathbb{I}+\dfrac{1}{2}\hat{\cal{R}}_{\a\n}\right)+\right.\nonumber\\&&+\left.g_{\a\m}\left(\dfrac{1}{6}R_{\b\n}\mathbb{I}+\dfrac{1}{2}\hat{\cal{R}}_{\b\n}\right)+g_{\a\b}\left(\dfrac{1}{6}R_{\m\n}\mathbb{I}+\dfrac{1}{2}\hat{\cal{R}}_{\m\n}\right)-\dfrac{1}{12}g^{(2)}_{\m\n\a\b}\mathbb{I}\right\}
\end{eqnarray}

\begin{eqnarray}
	\nabla_\a\nabla_\b\nabla_\m\nabla_\n\nabla_\s\nabla_\l\dfrac{\mathbb{I}}{\Box^3}&=& -\dfrac{\sqrt{g}}{8(n-4)\pi^2}\dfrac{1}{6}\left\{g^{(2)}_{\m\n\a\b}\hat{B}_{\s\l}+g^{(2)}_{\m\n\a\s}\hat{B}_{\b\l}+g^{(2)}_{\m\n\b\s}\hat{B}_{\a\l}+g^{(2)}_{\m\a\b\s}\hat{B}_{\n\l}\right.+\nonumber\\&&\left.+g^{(2)}_{\m\n\a\l}\hat{B}_{\b\s}+g^{(2)}_{\m\n\b\l}\hat{B}_{\a\s}+g^{(2)}_{\m\a\b\l}\hat{B}_{\n\s}+g^{(2)}_{\n\a\b\l}\hat{B}_{\m\s}+g^{(2)}_{\m\n\s\l}\hat{B}_{\a\b}\right.+\nonumber\\&&\left.+g^{(2)}_{\m\a\s\l}\hat{B}_{\n\b}+g^{(2)}_{\n\a\s\l}\hat{B}_{\m\b}+g^{(2)}_{\m\b\s\l}\hat{B}_{\n\a}+g^{(2)}_{\n\b\s\l}\hat{B}_{\m\a}+g^{(2)}_{\a\b\s\l}\hat{B}_{\m\n}\right.-\nonumber\\&&\left.\left.- \dfrac{1}{12}\bigg[g_{\s\l}\left(R_{\b\n\a\m}+R_{\a\n\b\m}\right)+g_{\b\l}\left(R_{\s\n\a\m}+R_{\a\n\s\m}\right)+g_{\a\l}\left(R_{\s\n\b\m}+R_{\b\n\s\m}\right)+\right.\right.\nonumber\\&&\left.\left.+g_{\n\l}\left(R_{\s\a\b\m}+R_{\b\a\s\m}\right)+g_{\m\l}\left(R_{\s\a\b\n}+R_{\b\a\s\n}\right)+g_{\b\s}\left(R_{\l\n\a\m}+R_{\a\n\l\m}\right)+\right.\right.\nonumber\\&&\left.\left.+g_{\a\s}\left(R_{\l\n\b\m}+R_{\b\n\l\m}\right)+g_{\m\s}\left(R_{\l\a\b\n}+R_{\b\a\l\n}\right)+g_{\a\b}\left(R_{\l\n\s\m}+R_{\s\n\l\m}\right)+\right.\right.\nonumber\\&&\left.\left.+g_{\n\b}\left(R_{\l\a\s\m}+R_{\s\a\l\m}\right)+g_{\m\b}\left(R_{\l\a\s\n}+R_{\s\a\l\n}\right)+g_{\n\a}\left(R_{\l\b\s\m}+R_{\s\b\l\m}\right)+\right.\right.\nonumber\\&&\left.\left.+g_{\m\a}\left(R_{\l\b\s\n}+R_{\s\b\l\n}\right)+g_{\m\n}\left(R_{\l\b\s\a}+R_{\s\b\l\a}\right)+\dfrac{1}{8}g^{(3)}_{\m\n\a\b\s\l} R\right]\mathbb{I}\right\}
\end{eqnarray}

where the field strength $\hat{\cal{R}}_{\m\n}$ defined as in \eqref{field_strength} and

\bea
&& g^{(0)}=1\nonumber\\
&& g^{(1)}_{\m\n}=g_{\m\n}\nonumber\\
&& g^{(2)}_{\m\n\a\b}=g_{\m\a}g_{\n\b}+g_{\m\b}g_{\n\a}+g_{\m\n}g_{\a\b}\nonumber\\
&& g^{(3)}_{\m\n\a\b\s\l}=g_{\m\n}g^{(2)}_{\a\b\s\l}+g_{\m\a}g^{(2)}_{\n\b\s\l}+g_{\m\b}g^{(2)}_{\n\a\s\l}+g_{\m\s}g^{(2)}_{\n\a\b\l}+g_{\m\l}g^{(2)}_{\n\a\b\s}\nonumber\\
&& g^{(n+1)}_{\m_1....\m_{2n+2}}=\sum_{i=2}^{2n+2}g_{\m_1\m_i}g^{(n)}_{\m_2...\m_{i-1}\m_{i+1}\m_{2n+2}}\nonumber\\
&& \hat{B}_{\a\b}=\dfrac{1}{24}R_{\a\b}\mathbb{I}+\dfrac{1}{8}\hat{\cal{R}}_{\a\b}
\eea 

Finally for p=2n-4 all traces can be computed with the expression

\begin{flalign}
&	\nabla_{\m_1}.....\nabla_{\m_{2n-4}}\dfrac{\mathbb{I}}{\Box^n}= -\dfrac{\sqrt{g}}{8(n-4)\pi^2}\dfrac{g^{(n-2)}_{\m_1....\m_{2n-4}}}{2^{n-2}(n-1)!}&
\end{flalign}

\newpage
\section{Quadratic divergences}\label{app_D}

When a proper time cutoff is introduced, quadratic divergences are characterized by the coefficient $a_2$ in a small time expansion of the heat kernel. This calculation is 
more or less standard for minimal operators, but some devious procedure is needed for the nonminimal part.

Actually, for quadratic minimal operators of the form
\be
D=-\gamma_{AB}\Box-E_{AB}
\ee
the \cite{DeWitt,Gusynin} $a_2$ coefficient  reads

\be
a_2=\dfrac{1}{6}\int d^4 x \sqrt{|g|}Tr(6E+R)\label{a2}
\ee
For quartic minimal operators
\begin{align}\label{quartic_operator}
	D=\gamma_{AB}\Box^{2}+\Omega_{AB}^{\m\n\a}\D_{\m}\D_{\n}\D_{\a}+J_{AB}^{\m\n}\D_{\m}\D_{\n}+H_{AB}^{\m}\D_{\m}+P_{AB}
\end{align}

and in case $\Omega^{\m\n\a}_{AB}=0$ (which is all we need here) the $a_2$ coefficient reads

\be
a_2=\sqrt{\pi}\int d^4 x \sqrt{|g|}\left(\dfrac{1}{6}R+\dfrac{1}{8}J^\m_\m\right)\label{a2_quartic}
\ee


The contribution of $S_{bc}$ to the quadratic divergence reads
\begin{align}
	a_2^{bc}=\dfrac{\sqrt{\pi}}{6}\int d^4 xR
\end{align}
where we have set $|g|=1$ and we have multiplied by minus two in order to take into account of the fact that there are two fermionic fields.
\bi

\item The contribution of $S_\pi$ reads
\be
a_2^\pi=\dfrac{1}{3}\int d^4 x R
\ee

where a minus two factor appears as  we are integrating over two fermionic fields.

\item The contribution of $S_{\bar{c}\phi}$ is
\be
a_2^{\bar{c}\phi}=\dfrac{1}{3}\int d^4 x R
\ee
there is now a factor of two that has been introduced to take into account that we have two fields.

\item The contribution of the Weyl ghost field  reads
\be
a_2^W=-\dfrac{1}{3}\int d^4 x R
\ee
where a factor of minus two has been introduced to take into account that we have two fermionic fields.

\item To summarize,  the total contribution of the ghost sector is

\be
a_2^{\text{gh}}=\dfrac{2+\sqrt{\pi}}{6}\int d^4 x R
\ee

\item In order to compute the contribution of the nonminimal piece, we shall employ  Toms' technique \cite{Toms} asserting that the coefficient $a_2$ for a quadratic operator $\hat{F}$ is just the pole part of the Green function in dimensional regularization. This yields

\be
a_2^{\text{non minimal}}=\dfrac{1}{6}\int d^4 x R
\ee

Adding all contributions the resulting coefficient reads

\be
a_2=\dfrac{3+\sqrt{\pi}}{6}\int d^4 x R
\ee

The gauge invariant and physical  divergence is obtained when it is on shell, that is, when
\be
R=C=\text{constant}
\ee
so that  quadratic divergences are analogous to the quartic ones in that they are non-dynamical, that is, they do not couple to the gravitational field.

\be
a_2^{\text{ on shell}}=C \dfrac{3+\sqrt{\pi}}{6}\int d^4 x 
\ee

\ei

\end{document}